\documentclass[%
aps,
prx,
reprint,
nofootinbib,
amsmath,
amssymb,
longbibliography
]{revtex4-2}
\usepackage{
physics,
graphicx,
multirow,
mathtools, 
placeins, 
nicefrac, 
hyperref 
}
\usepackage{adjustbox}
\usepackage[dvipsnames]{xcolor}
\hypersetup{
    colorlinks,
    linkcolor={blue!50!black},
    citecolor={blue!50!black},
    urlcolor={blue!80!black}
}

\usepackage[caption=false]{subfig} 
\usepackage[capitalize]{cleveref}

\newcommand{\supop}[1]{\ensuremath{\overbracket[0.1ex][0.3ex]{#1}}}
\renewcommand{\var}[1]{{\mathrm{Var}\left[#1\right]}}
\DeclarePairedDelimiter\pbra{\langle\!\langle}{\rvert}
\DeclarePairedDelimiter\pket{\lvert}{\rangle\!\rangle}
\DeclarePairedDelimiterX\pbraket[2]{\langle\!\langle}{\rangle\!\rangle}{#1 \delimsize\vert #2}

\begin{document}


\title{Multi-exponential Error Extrapolation and \\Combining Error Mitigation Techniques for NISQ Applications}


\author{Zhenyu Cai}
\email{cai.zhenyu.physics@gmail.com}
\affiliation{Department of Materials, University of Oxford, Oxford, OX1 3PH, United Kingdom}
\affiliation{Quantum Motion Technologies Ltd, Nexus, Discovery Way, Leeds, LS2 3AA, United Kingdom}

\date{\today}

\begin{abstract}
    Noise in quantum hardware remains the biggest roadblock for the implementation of quantum computers. To fight the noise in the practical application of near-term quantum computers, instead of relying on quantum error correction which requires large qubit overhead, we turn to quantum error mitigation, in which we make use of extra measurements. Error extrapolation is an error mitigation technique that has been successfully implemented experimentally. Numerical simulation and heuristic arguments have indicated that exponential curves are effective for extrapolation in the large circuit limit with an expected circuit error count around unity. In this Article, we extend this to multi-exponential error extrapolation and provide more rigorous proof for its effectiveness under Pauli noise. This is further validated via our numerical simulations, showing orders of magnitude improvements in the estimation accuracy over single-exponential extrapolation. Moreover, we develop methods to combine error extrapolation with two other error mitigation techniques: quasi-probability and symmetry verification, through exploiting features of these individual techniques. As shown in our simulation, our combined method can achieve low estimation bias with a sampling cost multiple times smaller than quasi-probability while without needing to be able to adjust the hardware error rate as required in canonical error extrapolation. 
\end{abstract}

\maketitle

\section{Introduction}
While fault-tolerant quantum computers promise huge speed-up over classical computers in areas like chemistry simulations, optimisation and decryption, their implementations remain a long term goal due to the large qubit overhead required for quantum error correction. With the recent rapid advance of quantum computer hardware in terms of both qubit quantity and quality, culminating with the ``quantum supremacy'' experiment~\cite{aruteQuantumSupremacyUsing2019}, one must wonder is it possible for us to perform classically intractable computations on such Noisy Intermediate-Scale Quantum (NISQ) hardware without quantum error correction~\cite{preskillQuantumComputingNISQ2018}. Resource estimation has been performed for one of the most promising applications on NISQ hardware: the Fermi-Hubbard model simulation~\cite{caiResourceEstimationQuantum2020, cadeStrategiesSolvingFermiHubbard2020}, realising that even with an optimistic local gate error rate of $10^{-4}$, the large number of gates needed for a classically intractable calculation will lead to an expected circuit error count of the order of unity. To obtain any meaningful results under such an expected circuit error count, it is essential to employ error mitigation techniques, which relies on making extra measurements, as opposed to employing extra qubits in the case of quantum error correction, to estimate the noise-free expectation values from the noisy measurement results. Three of the most well-studied error mitigation techniques are symmetry verification~\cite{mcardleErrorMitigatedDigitalQuantum2019, bonet-monroigLowcostErrorMitigation2018}, quasi-probability and error extrapolation~\cite{liEfficientVariationalQuantum2017, temmeErrorMitigationShortDepth2017, endoPracticalQuantumError2018}. 

Previously all of these error mitigation techniques have been discussed separately. They make use of different information about the hardware and the computation problems to perform different sets of extra circuit runs for error mitigation. Symmetry verification makes use of the symmetry in the simulated system and performs circuit runs with additional measurements. Quasi-probability makes use of the error models of the circuit components and performs circuit runs with different additional gates in the circuit. Error extrapolation makes use of the knowledge about tuning the noise strength via physical control of the hardware, and performs additional circuit runs at different noise levels. Consequently, the three error mitigation techniques are equipped to combat different types of noise with different additional sampling costs (number of additional circuit runs required). Hence, it is natural to wonder how these techniques might complement each other. For NISQ application, it may be essential to understand and develop ways for these error mitigation techniques to work in unison, to achieve better performance than the individual techniques in terms of lower bias in the noise-free expectation values estimates and/or lower sampling costs. Thus one key focus of our Article is on the methods for combining these error mitigation techniques, and trying to gauge their performance under different scenarios through analytical arguments and numerical simulations.

To achieve efficient combinations of these mitigation techniques, we will need to exploit certain features of these constituent techniques. As we will see later, we will show that quasi-probability can be used for error transformation instead of error removal and the circuit runs that fail the symmetry verification can actually be utilised instead of being discarded. It is also essential to understand the mechanism behind error extrapolation, especially in the NISQ limit in which the number of errors in the circuit will follow a Poisson distribution. Heuristic arguments and numerical validations have been made by Endo~\textit{et~al.}~\cite{endoPracticalQuantumError2018} on error extrapolation using exponential decay curves in this NISQ limit. However, it cannot be applied to certain situations arising in practice, for example when the data points have any increasing trend. Our Article will take this further and provide a more rigorous argument showing that single-exponential error extrapolation is just a special case of the more general multi-exponential error extrapolation framework, using which we can achieve a much lower estimation bias.

\section{Symmetry Verification}\label{sec:sym_proj}
Suppose we want to perform a state preparation and we know that the correct state must follow a certain symmetry $S$, i.e. we expect our end state to be the eigenstate of $S$ with the correct eigenvalue $s$ (or within a set of eigenvalues $\{s\}$). In such a case, we can perform $S$ measurement on our output state and discard the circuit runs that produce states that violate our symmetry. This was first proposed and studied by McArdle~\textit{et al.}~\cite{mcardleErrorMitigatedDigitalQuantum2019} and Bonet-Monroig~\textit{et al.}~\cite{bonet-monroigLowcostErrorMitigation2018}. Discarding erroneous circuit runs results in an effective density matrix that is the original density matrix $\rho$ projected into the $S = s$ subspace via the projection operator $\Pi_s$:
\begin{align*}
    \rho_s = \frac{\Pi_s\rho\Pi_s}{\Tr(\Pi_s \rho \Pi_s)} =  \frac{\Pi_s\rho\Pi_s}{\Tr(\Pi_s \rho)}. 
\end{align*}
Here we have used $\Pi_s \Pi_s = \Pi_s$. 

Now let us suppose we want to measure an observable $O$, which commutes with our symmetry $S$. Thus they can both be measured in the same run and we will discard the measurement results in the runs that failed the symmetry verification. The symmetry-verified expectation value of the observable $O$ is then:
\begin{align}\label{eqn:sym_proj}
    \expval{O_{sym}} = \Tr(O \rho_s)= \frac{\Tr(O\Pi_s\rho)}{\Tr(\Pi_s \rho)} \equiv \frac{\expval{O\Pi_s}}{\expval{\Pi_s}}
\end{align}
in which we have used $[S, O] = 0\ \Rightarrow\ [\Pi_s, O] = 0$. Note that $\Pi_s$ measurement takes the value $1$ if the symmetry verification is passed and $0$ otherwise and hence $\expval{\Pi_s} = \Tr(\Pi_s \rho)$ is just the fraction of circuit runs that fulfil the symmetry condition. We will use $P_d$ to denote the fraction of circuit runs that fail the symmetry verification, which gives
\begin{align*}
    \expval{\Pi_s} = \Tr(\Pi_s \rho) = 1 - P_d.
\end{align*}
Recall that $\rho_s$ is the effective density matrix of the non-discarded runs, which as mentioned is a $\Tr(\Pi_s \rho)$ fraction of the total number of runs. Therefore to obtain statistics from $\rho_s$, we need a factor of
\begin{align}\label{eqn:group_sym_cost_dir}
    C_{S} = \frac{1}{\Tr(\Pi_s \rho)} = \frac{1}{1 - P_d}
\end{align}
more circuit runs than obtaining directly from $\rho$. 

In the method discussed above, $O\Pi_s$ is usually obtained through measuring $O$ and $S$ in the same run.  However, sometimes this cannot be done due to, for example, inability to perform non-demolishing measurements. In such a case we need to break $O\Pi_s$ into its Pauli basis~\cite{bonet-monroigLowcostErrorMitigation2018} and reconstruct it via post-processing, this is discussed in \cref{sec:post_proc_proj}. In this Article, we will mainly be focusing on direct symmetry verification instead of post-processing verification, but most of the arguments are valid for both methods besides discussions about costs.

Now let us move on to see what errors are detectable by symmetry verification. We want to produce the state $\ket{\psi_f}$ which is known to fall within the $S = s$ symmetry subspace:
\begin{align*}
    S\ket{\psi_f} & = s \ket{\psi_f}.
\end{align*}
To produce the state, we usually start with a state $\ket{\psi_0}$ that follows the same symmetry and uses a circuit $U$ that consists of components that conserve the symmetry:
\begin{align*}
    \ket{\psi_f} & = U \ket{\psi_0}, \quad
    S\ket{\psi_0} = s\ket{\psi_0}, \quad \left[U, S\right] = 0.
\end{align*}
Suppose that some error $E$ occurs during the circuit in between the symmetry-preserving components and it satisfies
\begin{align}\label{eqn:cond_E}
    \Pi_sE &= E\Pi_{s'}
\end{align}
in which $s, s'$ are some eigenvalues of the symmetry operator $S$. We then have:
\begin{align}
    \Pi_sE\Pi_s &= E\Pi_{s'}\Pi_s = \begin{cases}
        E \Pi_s \quad &s = s'\\
        0 \quad &s \neq s'\label{eqn:cond_E_2}
    \end{cases}
\end{align}
$\Pi_s E\Pi_s = E\Pi_s$ means that $E$ is a transformation within the $S = s$ subspace, hence $E$ is undetectable by the symmetry verification using $S$. $\Pi_s E\Pi_s = 0$ means that  $E$ contains no components that map states in the $S = s$ subspace back into the same subspace, hence $E$ is (completely) detectable by the symmetry verification using $S$. A general error will be a combination of detectable and undetectable error components.

In this Article, we will be focusing on Pauli errors and Pauli symmetries, for which \cref{eqn:cond_E_2} is reduced to:
\begin{equation}\label{eqn:pauli_sym_ver}
    \begin{aligned}
        \left[S, E\right] &= 0 \Rightarrow s = s' \quad \quad E\text{ is undetectable}\\
        \left\{S, E\right\} &= 0 \Rightarrow s = -s' \quad \quad E\text{ is detectable}.
    \end{aligned}
\end{equation}

\section{Quasi-probability}\label{sec:qua_prob}
To describe the quasi-probability method, we will make use of the Pauli transfer matrix (PTM) formalism~\cite{greenbaumIntroductionQuantumGate2015}. Using $\mathbb{G}$ to denote the set of Pauli operators, any density operators can be written in the vector form by decomposing into the Pauli basis $G \in \mathbb{G}$:
\begin{align*}
    \rho &= \frac{1}{2^N}\sum_{G \in \mathbb{G}} \Tr(G\rho) G \\
    \Rightarrow \pket{\rho} &= \sum_{G \in \mathbb{G}} \pket{G}\pbraket{G}{\rho}
\end{align*}
where we have defined the inner product as:
\begin{align*}
    \pbraket{G}{\rho}  = \frac{1}{\sqrt{2^N}} \Tr(G\rho).
\end{align*}
Note that we need to add a normalisation factor $\nicefrac{1}{\sqrt{2^N}}$ when we use the Pauli operators as basis, with $N$ being the number of qubits. 

The quasi-probability method was first introduced by Temme~\textit{et.al.}~\cite{temmeErrorMitigationShortDepth2017}, and the implementation details were later studied by Endo~\textit{et.al.}~\cite{endoPracticalQuantumError2018}. Let us suppose we are trying to perform the operation $\mathcal{U}$, but in practice we can only implement its noisy version
\begin{align*}
    \mathcal{U}_{\epsilon} = \mathcal{E}\mathcal{U}.
\end{align*}
In addition to $\mathcal{U}_{\epsilon}$, we can also implement a set of basis operation $\{\mathcal{B}_n\}$. We can decompose the ideal operation $\mathcal{U}$ that we \emph{want to} implement into a set of gates $\{\mathcal{B}_n\mathcal{U}_{\epsilon}\}$ that we \emph{can} implement:
\begin{align*}
    \mathcal{U} &= \sum_{n} q_n \mathcal{B}_n\mathcal{U}_{\epsilon} \quad \Rightarrow \quad \mathcal{E}^{-1} = \sum_{n} q_n \mathcal{B}_n.
\end{align*}
In this way, we are essentially trying to simulate the behaviour of the inverse noise channel $\mathcal{E}^{-1}$ using the set of basis operations $\{\mathcal{B}_n\}$, which can undo the noise $\mathcal{E}$. 

If we have a state $ \pket{\rho}$ passing through the circuit $\mathcal{U}$ and we perform measurement $O$, then the observable we obtain during the experiment will be:
\begin{align*}
    \expval{O} = \pbra{O} \mathcal{U} \pket{\rho} &= \sum_{n} q_n \pbra{O} \mathcal{B}_n \mathcal{U}_{\epsilon}\pket{\rho}\\
    & = Q \sum_{n} s_n \frac{\abs{q_n}}{Q} \pbra{O} \mathcal{B}_n \mathcal{U}_{\epsilon}\pket{\rho}.
\end{align*}
in which $Q  = \sum_{n} \abs{q_n}$ and $s_n = \text{sgn}(q_n)$. This is implemented by sampling from the set of basis operations $\{\mathcal{B}_n\}$ with the probability distribution $\{\frac{\abs{q_n}}{Q}\}$. We will weight each measurement outcome by the sign factor $s_n$ and rescale the final expectation value by the factor $Q$.

Now if we break down our computation into many components $\mathcal{U} = \prod_{m = 1}^{M}\mathcal{U}_{m}$, with noise associated with each component, then the observable that we want to measure is:
\begin{align*}
    \expval{O} = \pbra{O} \prod_{m = 1}^{M}\mathcal{U}_{m} \pket{\rho},
\end{align*}
but in reality we can only implement the noisy version:
\begin{align*}
    \expval{O_{\epsilon}} = \pbra{O} \prod_{m = 1}^{M}\mathcal{E}_{m}\mathcal{U}_{m} \pket{\rho}.
\end{align*}
Each noise element can be removed by simulating the inverse channels using the set of basis operations $\{\mathcal{B}_n\}$:
\begin{align*}
    \mathcal{E}_{m}^{-1} = \sum_{n} q_{mn} \mathcal{B}_{n} = R_m \sum_{n} s_{mn} \frac{\abs{q_{mn}}}{R_m} \mathcal{B}_{n}
\end{align*}
with $R_m = \sum_{n} \abs{q_{mn}}$.

Hence, we can get back the noiseless observable using
\begin{align*}
    \expval{O} &=   \pbra{O} \prod_{m = 1}^{M} \left(\sum_{n_m} q_{mn_m} \mathcal{B}_{n_m}\right)\mathcal{E}_{m}\mathcal{U}_{m} \pket{\rho}\\
    & = Q\sum_{\vec{n}} s_{\vec{n}} \frac{\abs{q_{\vec{n}}}}{Q} \pbra{O} \prod_{m = 1}^{M} \mathcal{B}_{n_m}\mathcal{E}_{m}\mathcal{U}_{m} \pket{\rho}
\end{align*}
in which we have used $\vec{n}$ to denote the set of number $\{n_1, n_2, \cdots, n_M\}$ and we have defined $q_{\vec{n}} = \prod_{m = 1}^{M} q_{mn_m}$, 
$s_{\vec{n}}  = \prod_{m = 1}^{M} s_{mn_m} = \text{sgn}(q_{\vec{n}})$ and 
\begin{align}\label{eqn:cost_Q}
    Q  = \prod_{m = 1}^{M} R_m = \prod_{m = 1}^{M} \sum_{n_m} \abs{q_{mn_m}}.
\end{align}
To implement this, we simply sample the set of basis operations $\{\mathcal{B}_{n_1}, \mathcal{B}_{n_2}, \cdots, \mathcal{B}_{n_M}\}$ that we want to implement with the probability $\nicefrac{\abs{q_{\vec{n}}}}{Q}$ and weight each measurement outcome by a sign factor $s_{\vec{n}} = \text{sgn}(q_{\vec{n}})$, so that the outcome we get is an effective Pauli observable $O_Q$. And the error-free observable expectation value can be obtained via:
\begin{align}\label{eqn:qus_trans_var}
    \expval{O} = Q \expval{O_Q}.
\end{align}
Hence, to estimate $\expval{O}$ by sampling $O_Q$, we need $C_{Q}$ times more samples than sampling $O$ directly, where the sampling cost factor $C_{Q}$ is:
\begin{align}\label{eqn:quasi_cost}
    C_{Q} = Q^2 = \left(\prod_{m = 1}^{M} \sum_{n_m} \abs{q_{mn_m}}\right)^2.
\end{align}

In this Article we will be mainly focusing on \emph{Pauli error channels}, which can be inverted using quasi-probability by employing Pauli gates as the basis operations. 

Using $\supop{\quad}$ to denote a super-operator
\begin{align*}
    \left(\supop{A} + \supop{B}\right) \rho = A \rho A^\dagger + B \rho B^\dagger,
\end{align*}
any Pauli channel can be written in the form:
\begin{align}\label{eqn:qua_example_pauli}
    \mathcal{G}_{p_\epsilon} = (1 - p_\epsilon) \mathcal{I} + p_\epsilon \sum_{G \in \mathbb{G} - I} \alpha_G \supop{G}
\end{align}
with $\sum_G \alpha_G = 1$. We use $p_\epsilon$ to denote the total probability of all the \emph{non-identity} components. This channel can be \emph{approximately} inverted using the quasi-probability channel $\mathcal{G}_{-p_\epsilon}$ since:
\begin{align*}
    \mathcal{G}_{-p_\epsilon} \mathcal{G}_{p_\epsilon} \approx \mathcal{I} + \mathcal{O}(p_\epsilon^2).
\end{align*}
Hence, to the first order approximation, the cost of inverting $\mathcal{G}_{p_\epsilon}$ will be the cost of implementing $\mathcal{G}_{-p_\epsilon}$, which using \cref{eqn:quasi_cost} is
\begin{align}\label{eqn:cost_grp_qua_2}
    C_{Q1, 0} \approx (1 + 2p_\epsilon)^2 \approx 1 + 4p_\epsilon.
\end{align}

Here we have only discussed approximately inverting a Pauli channel because the exact inverse channel can be hard to express in a compact analytical form. However, it can be obtained numerically by first obtaining the Pauli transfer matrix of the noise channel and then performing matrix inversion.

Instead of removing the error channel completely, quasi-probability can also be used to transform the form of an error channel. In the case of Pauli channels, suppose we want to transform a channel of the form in \cref{eqn:qua_example_pauli} to
\begin{align*}
    \mathcal{F}_{q_\epsilon} = (1 - q_\epsilon) \mathcal{I} + q_\epsilon \sum_{G \in \mathbb{G} - I} \beta_G \supop{G},
\end{align*}
we can \emph{approximately} achieve this transformation up to first order in $q_{\epsilon}$ and $p_{\epsilon}$ by applying the quasi-probability channel:
\begin{align*}
    \mathcal{R} = (1 + \left(p_\epsilon - q_\epsilon\right)) \mathcal{I} - \sum_{G \in \mathbb{G} - I} \left(p_\epsilon \alpha_G - q_\epsilon \beta_G\right) \supop{G}.
\end{align*}
This will incur the implementation cost:
\begin{align*}
    C_{Q1, q} &= \left(\abs{1 + \left(p_\epsilon - q_\epsilon\right)} + \sum_{G \in \mathbb{G} - I} \abs{p_\epsilon \alpha_G - q_\epsilon \beta_G}\right)^2.
\end{align*}
In the limit of small $p_\epsilon$ and $q_\epsilon$, we have:
\begin{align*}
    C_{Q1, q} & \approx 1 + 2\left(p_\epsilon - q_\epsilon\right) + 2\sum_{G \in \mathbb{G} - I} \abs{p_\epsilon \alpha_G - q_\epsilon \beta_G}\\
    & = 1 + 4\sum_{\substack{\tiny G \in \mathbb{G} - I,\\ p_\epsilon \alpha_G > q_\epsilon \beta_G}} \left(p_\epsilon \alpha_G - q_\epsilon \beta_G\right).
\end{align*}
In the last step we have used $\sum_{G} \left(p_\epsilon \alpha_G - q_\epsilon \beta_G\right) = p_\epsilon - q_\epsilon$ from $\sum_G \alpha_G = \sum_G \beta_G = 1$.

If we are suppressing all error components evenly, or if we are simply removing certain error components, we will have $p_\epsilon \alpha_G \geq q_\epsilon \beta_G \quad \forall G \in \mathbb{G} - I$. In this case, the cost of implementing the transformation using quasi-probability will simply be:
\begin{align}\label{eqn:pauli_trans_cost}
    C_{Q1, q} \approx 1+4\left(p_\epsilon - q_\epsilon\right).
\end{align}

\section{Group Errors}\label{sec:group_channel}
Here we will introduce a special kind of error channel: \emph{group error channels}, which enable us to make more analytical predictions about the error mitigation techniques that we have already discussed and also will help our understanding about error extrapolation later.

The group error $\mathcal{J}_{p, \mathbb{E}}$ of the group $\mathbb{E}$ is defined to be:
\begin{align}\label{eqn:err_chan}
    \mathcal{J}_{p, \mathbb{E}} &= (1- p) \supop{I} + \frac{p}{\abs{\mathbb{E}}} \sum_{E \in \mathbb{E}} \supop{E}
\end{align}
By groups we mean the subgroups of the Pauli group with a composition rule that ignores all the irrelevant phase factors. For the case of $p=1$, we will call $\mathcal{J}_{1, \mathbb{E}}$ the \emph{pure} group errors. 

Many physically interesting noise models like depolarising channels, dephasing channels, Pauli-twirled swap errors and dipole-dipole errors are all group errors.

Now let us consider the effect of applying a set of Pauli symmetry checks $\mathbb{S}$ to the group error in \cref{eqn:err_chan}. Using \cref{eqn:pauli_sym_ver}, $\mathbb{S}$ can remove and detect components in $\mathcal{J}_{p, \mathbb{E}}$ that anti-commute with any elements in $S \in \mathbb{S}$. We look at the action of $\mathbb{S}$ on the subset of qubits affected by $\mathcal{J}_{p, \mathbb{E}}$, and denote the set of these operators on the subset of qubits as $\mathbb{S}_{sub}$. The commutation relationship between $\mathbb{S}$ and $\mathbb{E}$ is equivalent to that of $\mathbb{S}_{sub}$ and $\mathbb{E}$. We denote their generators as $\widetilde{\mathbb{S}}_{sub}$ and $\widetilde{\mathbb{E}}$. Note that here $\mathbb{S}_{sub}$ is not a group, by $\widetilde{\mathbb{S}}_{sub}$ we just mean the set of independent elements in $\mathbb{S}_{sub}$. For Pauli generators, we can choose $\widetilde{\mathbb{E}}$ in such a way that for every $\widetilde{S}_{sub} \in \widetilde{\mathbb{S}}_{sub}$, there will at most be only one element in $\widetilde{\mathbb{E}}$ that anti-commutes with it. We will denote the elements in $\widetilde{\mathbb{E}}$ that commute with all elements in $\widetilde{\mathbb{S}}_{sub}$ as $\widetilde{\mathbb{Q}}$:
\begin{align*}
    \widetilde{\mathbb{Q}} = \{\widetilde{E} \in \widetilde{\mathbb{E}}\ |\ \left[\widetilde{E}, \widetilde{S}_{sub}\right] = 0 \quad \forall  \widetilde{S}_{sub} \in \widetilde{\mathbb{S}}_{sub}\}
\end{align*}
and it will generate the remaining error components in $\mathbb{E}$ that are not detectable, which we denote as $\mathbb{Q}$. Hence the detectable error components are just $\mathbb{E} - \mathbb{Q}$

Going back to our error channel in \cref{eqn:err_chan}, the probability that the error gets detected is just the total probability of the detectable error components:
\begin{align}\label{eqn:prob_detected}
    p_d = \frac{\abs{\mathbb{E}} - \abs{\mathbb{Q}}}{\abs{\mathbb{E}}}p.
\end{align}
Removing the detected errors in $\mathcal{J}_{p, \mathbb{E}}$ and renormalising the error channel by the factor $1 - p_d$ gives the effective channel after verification, which is just another group error channel
\begin{align}\label{eqn:resultant_group_ch}
    \mathcal{J}_{r, \mathbb{Q}} &= (1- r) \supop{I} + \frac{r}{\abs{\mathbb{Q}}} \sum_{Q \in \mathbb{Q}} \supop{Q}
\end{align}
with
\begin{align*}
    r = \frac{\abs{\mathbb{Q}}p}{\abs{\mathbb{E}}\left(1 - p_d\right)}=  \frac{\abs{\mathbb{Q}}p}{\abs{\mathbb{E}} \left(1-p\right)  + \abs{\mathbb{Q}}p} \approx  \frac{\abs{\mathbb{Q}}}{\abs{\mathbb{E}}} p + \mathcal{O}(p^2).
\end{align*}
An example of removing detectable errors for depolarising channels will be worked out later in \cref{sec:expec_decay_sim}.

For a given general Pauli channel, we have only discussed its \emph{approximate} inverse channel in \cref{sec:qua_prob}. This is because its \emph{exact} inverse channel can be hard to express in a compact analytical form. However, for any group channel, we can easily write down the explicit form of its inverse channel. 

As shown in \cref{sec:prop_group_channel}, it is easy to verify that the inverse of a group channel $\mathcal{J}_{p, \mathbb{E}}$ is just:
\begin{align}\label{eqn:group_inverse}
    \left(\mathcal{J}_{p, \mathbb{E}}\right)^{-1} = \mathcal{J}_{-\alpha, \mathbb{E}} 
    & = \left(1+ \alpha\right) \mathcal{I} - \alpha\mathcal{J}_{1, \mathbb{E}}
\end{align}
with $\alpha = \frac{p}{1 - p}$.

Using \cref{eqn:group_inverse} and \cref{eqn:quasi_cost}, the cost of using quasi-probability to invert $\mathcal{J}_{p, \mathbb{E}}$ is thus:
\begin{align}\label{eqn:cost_grp_qua}
    C_{Q1, 0} 
    &= \left(1 + 2\frac{\left(\abs{\mathbb{E}} - 1\right)p}{\abs{\mathbb{E}}(1-p)}\right)^2 \nonumber\\
    & \approx 1 + 4\frac{\abs{\mathbb{E}} - 1}{\abs{\mathbb{E}}}p + \mathcal{O}(p^2),
\end{align}
which is the same as \cref{eqn:cost_grp_qua_2} with
\begin{align}\label{eqn:p_e}
    p_{\epsilon} = \frac{\abs{\mathbb{E}} - 1}{\abs{\mathbb{E}}}p.
\end{align}

As shown in \cref{eqn:resultant_group_ch}, for a given group error $\mathcal{J}_{p, \mathbb{E}}$, the resultant error channel after symmetry verification is another group channel $\mathcal{J}_{r, \mathbb{Q}}$ where $\mathbb{Q}$ is a subgroup of $\mathbb{E}$ and $r = \frac{\abs{\mathbb{Q}}}{\abs{\mathbb{E}}}p$. The remaining errors can then be completely removed by implementing $\mathcal{J}_{r, \mathbb{Q}}^{-1}$ using quasi-probability. 

Similarly if we implement the quasi-probability inverse channel $\mathcal{J}_{r, \mathbb{Q}}^{-1}$ first and then perform symmetry verification, we can still completely remove the group error $\mathcal{J}_{p, \mathbb{E}}$. The gates we need to implement in the inverse channel $\mathcal{J}_{r, \mathbb{Q}}^{-1}$ will not be detected and thus will not be affected by the symmetry verification. As shown in \cref{sec:channel_after_qua}, the resultant error channel after applying $\mathcal{J}_{r, \mathbb{Q}}^{-1}$ is:
\begin{align}\label{eqn:err_after_qua}
    \mathcal{J}_{r, \mathbb{Q}}^{-1} \mathcal{J}_{p, \mathbb{E}} & = \left(1-p_d\right) \supop{I}  + \frac{p_d}{\abs{\mathbb{E}} - \abs{\mathbb{Q}}} \sum_{V \in \mathbb{E}, V \not \in \mathbb{Q}} \supop{V}.
\end{align}
This is a channel that only contains the error components that are detectable by the symmetry verification with the error rate $p_d$.

As discussed in \cref{sec:qua_prob} and explicitly shown in \cref{sec:channel_after_qua}, we can implement additional quasi-probability operations to further reduce the error rate of the resultant channel to $q \leq p_d$. The resultant detectable error channel will be:
\begin{align}
    \mathcal{V}_q &= \left(1-q\right) \supop{I}  + \frac{q}{\abs{\mathbb{E}} - \abs{\mathbb{Q}}} \sum_{V \in \mathbb{E}, V \not \in \mathbb{Q}} \supop{V}\label{eqn:quasi_final_arbit_channel}\\
    & = \left(1-q\right) \supop{I}  + q \mathcal{V}_1 \quad \quad q\leq p_d\nonumber.
\end{align}

\section{NISQ Limit}\label{sec:nisq_limit}
The number of possible error locations in the circuit, which is usually the number of gates in the circuit, will be denoted as $M$. These error locations might experience different noise with different error rates. From here on, instead of building our discussions around local gate error rates, we will see that the more natural quantity to consider in the context of NISQ error mitigation is the expected number of errors occurring in each circuit run, which is called the \emph{mean circuit error count} and denoted as $\mu$. In order to achieve quantum advantage using NISQ machine, we would generally expect the circuit size to be large enough to be classically intractable while the mean circuit error count should be not be too far beyond unity:
\begin{equation}\label{eqn:poisson_limit}
    \begin{aligned}
        M \gg 1,\quad \mu \sim 1.
    \end{aligned}
\end{equation}
This is also called the Poisson limit since using Le Cam's theorem, the number of errors occurring in each circuit run will follow the Poisson distribution -- i.e. the probability that $l$ errors occur will be:
\begin{align}\label{eqn:poisson_prob}
    P_l = e^{-\mu}\frac{\mu^l}{l!}.
\end{align}

If we assume that every local error channel in the circuit can be approximated as the composition of an undetectable error channel and a detectable error channel, then symmetry verification will have no effects on the undetectable error channels and we can focus only on the detectable error channels. Alternatively, as we have seen in \cref{sec:group_channel}, we can use quasi-probability to remove all the local undetectable errors in the circuit, leaving us with only detectable error channels. The expected number of \emph{detectable} errors occurring in each circuit run is denoted as $\mu_d$. Taking the NISQ limit and using \cref{eqn:poisson_limit}, the probability that $l$ detectable errors occur in the circuit is:
\begin{align*}
    P_l = e^{-\mu_d}\frac{\mu_d^l}{l!}.
\end{align*}

Using \cref{eqn:pauli_sym_ver}, an odd number of detectable errors will anti-commute with the symmetry and get detected while an even number of detectable errors will commute with the symmetry and pass the verification. 
Therefore, the total probability that the errors in the circuit will be detected by the verification of \emph{one} Pauli symmetry is thus:
\begin{align}\label{eqn:nisq_detectable_prob}
    P_{d} = \sum_{\text{odd } l} P_l = e^{-\mu_d} \sinh(\mu_d) = \frac{1 - e^{-2\mu_d}}{2}.
\end{align}
Note that this is upper-bounded by $\frac{1}{2}$, i.e. at most we can catch errors in half of the circuit runs. 

Hence, using \cref{eqn:group_sym_cost_dir}, the cost of implementing symmetry verification for \emph{one} Pauli symmetry is:
\begin{align}\label{eqn:nisq_sym_cost}
    C_{S, \mu_d} = \frac{1}{1 - P_d}  = \frac{1}{e^{-\mu_d} \cosh(\mu_d)} = \frac{2}{1 + e^{-2\mu_d}}
\end{align}
which is upper-bounded by $2$. 

After symmetry verification, the fraction of circuit runs that still have errors inside is:
\begin{align}\label{eqn:circ_err_after_sym}
    P_{circ} = 1 - P_d - P_0 = \frac{1}{2} \left(1 - e^{-\mu_d}\right)^2.
\end{align}

In \cref{eqn:pauli_trans_cost} we have only been focusing on applying quasi-probability to one error channel. Assuming there are $M$ such channels in the circuit, then using \cref{eqn:pauli_trans_cost} and taking the NISQ limit (\cref{eqn:poisson_limit}), the sampling cost factor of transforming all $M$ error channels with error probability $p_\epsilon$ into new error channels with error probability $q_\epsilon \leq p_\epsilon$ using quasi-probability is:
\begin{align}\label{eqn:partial_qua_cost_Mq}
    C_{Q, Mq} = C_{Q1, q}^M \approx \lim\limits_{M \rightarrow \infty}\left(1 + 4\left(p_\epsilon - q_\epsilon\right)\right)^M = e^{4M\left(p_\epsilon - q_{\epsilon}\right)}.
\end{align}
At $q_\epsilon = 0$, we will obtain the sampling cost of removing all the noise using quasi-probability:
\begin{align}\label{eqn:pure_qua_cost}
    C_{Q, 0} \approx e^{4Mp_\epsilon}.
\end{align}

Remember that we are focusing on Pauli errors and $p_{\epsilon}$ is defined to be the total probability of all non-identity components. This is not equivalent to the error probability $p$ because sometimes there are some identity components in our definition of error probability such as the group errors that we discussed in \cref{sec:group_channel}. Similar to the definition of $p_{\epsilon}$, we can denote the expected number of \emph{non-identity} errors in each circuit run as $\mu_\epsilon$. In the above cases, we have $\mu_\epsilon = Mp_\epsilon$ and similarly we can define $\nu_\epsilon = Mq_\epsilon$. In the cases where different noise locations experience different noise, using Le Cam's theorem with negative probabilities and focusing on the zero-occurrence case, we can generalise \cref{eqn:partial_qua_cost_Mq} to:
\begin{align}\label{eqn:partial_qua_cost}
    C_{Q, \nu} = e^{4(\mu_\epsilon - \nu_\epsilon)}.
\end{align}
i.e. the sampling cost of quasi-probability transformation grows exponentially with the reduction in the error rate $\mu_\epsilon - \nu_\epsilon$.

\section{Multi-exponential Error Extrapolation}\label{sec:err_extrapolation}
The idea of amplifying the hardware error rate and performing extrapolation using the original result and the noise-amplified result was first introduced by  Li~\textit{et~al.}~\cite{liEfficientVariationalQuantum2017} and Temme~\textit{et~al.}~\cite{temmeErrorMitigationShortDepth2017}, and was later successfully realised experimentally using superconducting qubits~\cite{kandalaErrorMitigationExtends2019}. Endo~\textit{et~al.}~\cite{endoPracticalQuantumError2018} provided heuristic arguments on why the exponential decay curve should be used for error extrapolation in the large circuit limit, whose improved performance over linear extrapolation was validated via numerical simulations in Ref.~\cite{endoPracticalQuantumError2018, giurgica-tironDigitalZeroNoise2020a}.

Using $\mathbb{L}$ to denote the set of locations that the errors have occurred, when $l$ errors have occurred in the circuit, our observable $O$ will be transformed to a noisy observable $O_{\abs{\mathbb{L}} = l}$. Recall that in the NISQ limit, the probability that $l$ errors happen (denoted as $P_l$) will follows the Poisson distribution in \cref{eqn:poisson_prob}. Therefore, the expectation value of the observable $O$ at the mean circuit error count $\mu$ is then:
\begin{align}\label{eqn:err_obs_expand}
    \expval{O_{\mu}} & = \sum_{l = 0}^{\infty} P_l \expval{O_{\abs{\mathbb{L}} = l}} = e^{-\mu} \sum_{l = 0}^{\infty} \frac{\mu^l}{l!} \expval{O_{\abs{\mathbb{L}} = l}}.
\end{align}
Hence, how $\expval{O_{\mu}}$ changes with $\mu$ is entirely determined by how $\expval{O_{\abs{\mathbb{L}} = l}}$ changes with $l$. When we try to fit a $n$th degree polynomial of $\mu$ to $\expval{O_{\mu}}$, for example performing a linear extrapolation~\cite{liEfficientVariationalQuantum2017, temmeErrorMitigationShortDepth2017}, we are essentially assuming that $\expval{O_{\abs{\mathbb{L}} = l}}$ is a $n$th degree polynomial of $l$ using the expressions of the moments of the Poisson distribution. 

At $l = 0$, we have the error-free result $\expval{O}$:
\begin{align*}
    \expval{O_{\abs{\mathbb{L}} = 0}} &= \expval{O}.
\end{align*}
At large error number $l$, in the case of stochastic errors, the circuit will move closer to a random circuit. Hence, for a Pauli observable $O$ we will expect:
\begin{align*}
    \lim\limits_{l \rightarrow \infty}\expval{O_{\abs{\mathbb{L}} = l}} &= 0.
\end{align*}
A generic polynomial of $l$ will not satisfy the above boundary conditions. Hence, to align with the above boundary conditions, we can instead assume an exponential decay of $\expval{O_{\abs{\mathbb{L}} = l}}$ as $l$ increase:
\begin{align*}
    \expval{O_{\abs{\mathbb{L}} = l}} = \expval{O}(1 - \gamma)^{l}
\end{align*}
in which $\gamma$ is the observable decay rate that satisfies $0 \leq \gamma \leq 1$. This will lead to an exponential function in $\mu$:
\begin{align}
    \expval{O_{\mu}} & = \expval{O}e^{-\mu} \sum_{l = 0}^{\infty} \frac{\left(\mu (1- \gamma)\right)^l}{l!} = \expval{O}e^{-\gamma \mu},\label{eqn:r_to_p}
\end{align}
which is just the extrapolation curves employed in Ref.~\cite{endoPracticalQuantumError2018}.

Using \cref{eqn:r_to_p}, if we probe at the error rates $\mu$ and $\lambda \mu$, we can perform two-point exponential extrapolation and obtain the error-mitigated estimate of $\expval{O}$, denoted as $\expval{O_0}$, using the following equation:
\begin{align*}
    \expval{O_0} = \left(\frac{\expval{O_{\mu}}^\lambda}{\expval{O_{\lambda \mu}} }\right)^{\frac{1}{\lambda - 1}}.
\end{align*}
As discussed in \cref{sec:extr_cost_detail}, the sampling cost factor of performing such an extrapolation is
\begin{align}\label{eqn:extrapolate_exp_cost}
    C_E 
    &\approx 2\frac{\lambda^2 e^{2\gamma \mu} + e^{2\lambda \gamma  \mu}}{\left(\lambda - 1\right)^2}.
\end{align}

Now let us try to gain a deeper insight about the reason behind the exponential decay of $\expval{O_{\abs{\mathbb{L}} = l}}$. If a Pauli error $G$ occurs at the end of a circuit and we are trying to measure an Pauli observable $O$, then the expectation value is just:
\begin{align*}
    \Tr(G\rho G O) = \eta(G, O)\Tr(\rho O)
\end{align*}
where $\eta(G, O)$ is the commutator between $G$ and $O$:
\begin{align*}
    GO = \eta(G, O) OG.
\end{align*}

If a pure group error $\mathcal{J}_{1, \mathbb{E}}$ occurs at the end of the circuit, then the resultant expectation value is:
\begin{align*}
    \Tr(\mathcal{J}_{1, \mathbb{E}}(\rho) O) & = \frac{1}{\abs{\mathbb{E}}}\sum_{E \in \mathbb{E}} \Tr(E\rho E O)\\
    & = \left( \frac{1}{\abs{\mathbb{E}}}\sum_{E \in \mathbb{E}} \eta(E, O)\right) \Tr(\rho O)
\end{align*}
Using the fact that the composition of the commutators of elements in a Pauli subgroup follows the same structure as the composition of the elements themselves:
\begin{align*}
    \eta(EE', O) = \eta(E', O)\eta(E, O),
\end{align*}
we can rewrite the above formula in terms of the generator of $\mathbb{E}$:
\begin{align}\label{eqn:effect_group_ch}
    \Tr(\mathcal{J}_{1, \mathbb{E}}(\rho) O) 
    & = \left(\prod_{\widetilde{E} \in \widetilde{\mathbb{E}}} \frac{1 + \eta(\widetilde{E}, O)}{2}\right) \Tr(\rho O)
\end{align}
in which
\begin{align*}
    \prod_{\widetilde{E} \in \widetilde{\mathbb{E}}} \frac{1 + \eta(\widetilde{E}, O)}{2} & = \begin{cases}
        1 &\text{ if $\eta(E, O) = 1 \quad \forall E \in \mathbb{E}$}\\
        0 &\text{ otherwise}.
    \end{cases}
\end{align*}
Hence, if a pure group error $\mathcal{J}_{1, \mathbb{E}}$ occurs right before measuring a Pauli observable $O$, then the resultant expectation value $\Tr(\mathcal{J}_{1, \mathbb{E}}(\rho) O)$ will only remain unchanged if $O$ commutes with \emph{all} elements in $\mathbb{E}$, otherwise the information about $O$ is ``erased'' by the group error and the expectation value will be $0$.

If we decompose the gates in a unitary circuit $U = \prod_{m = M}^{1} V_m$ into their Pauli components: $V_m = \sum_{j_{m}} \alpha_{mj_{m}} G_{mj_m}$, then we have:
\begin{align*}
    U &= \prod_{m = M}^{1} \sum_{j_{m}} \alpha_{mj_{m}} G_{mj_m} = \sum_{\vec{j}}  \alpha_{\vec{j}} G_{\vec{j}}
\end{align*}
where
\begin{align*}
    \sum_{\vec{j}} = \prod_{m = M}^1 \sum_{j_m},\ G_{\vec{j}} = \prod_{m = M}^{1}  G_{mj_m},\ \alpha_{\vec{j}} = \prod_{m = M}^{1} \alpha_{mj_{m}}.
\end{align*}
i.e. the circuit $U$ can be viewed as the superposition of many Pauli circuits $G_{\vec{j}}$.

The expectation value of observing $O$ after applying the circuit $U$ on $\rho$ is 
\begin{align}\label{eqn:correct_exp}
    \expval{O} &= \Tr(U\rho U^\dagger O) \nonumber\\
    &= 2\sum_{\vec{i} > \vec{j}}  \Re{\alpha_{\vec{i}}^* \alpha_{\vec{j}}  \Tr(\rho G_{\vec{i}}^\dagger OG_{\vec{j}})}
\end{align}
which is a linear sum of the measurement results for the set of effective Pauli observables $G_{\vec{i}}^\dagger OG_{\vec{j}}$ for different $\vec{i}$ and $\vec{j}$. Similar to \cref{eqn:effect_group_ch}, the information about $G_{\vec{i}}^\dagger OG_{\vec{j}}$ will either be ``erased'' or perfectly preserved if a group error occurs in the circuit. Assuming the average fraction of group error locations in our circuit that can erase the information about $G_{\vec{i}}^\dagger OG_{\vec{j}}$ is $\gamma_{\vec{i}, \vec{j}}$, then as proven in \cref{sec:der_exp_decay},
in the limit of large $M$ and non-vanishing $1-\gamma_{\vec{i}, \vec{j}}$, the expectation value given $l$ errors occurred can be approximated to be:
\begin{align}
    \expval{O_{\abs{\mathbb{L}} = l}} &\approx  2\sum_{\vec{i} > \vec{j}}  \Re{\alpha_{\vec{i}}^* \alpha_{\vec{j}} \Tr(\rho G_{\vec{i}}^\dagger OG_{\vec{j}})} \left(1-\gamma_{\vec{i}, \vec{j}}\right)^l\label{eqn:OEl_exp},
\end{align}
which is just a multi-exponential decay curve.

If we consider the case in which many error locations in the circuit are affected by the same type of noise, and adding onto the fact that in practice there are usually many repetitions of the circuit structures along the circuit and across the qubits, we can expect many $\gamma_{\vec{i}, \vec{j}}$ of different $\vec{i}$ and $\vec{j}$ to be very similar. Hence, by grouping  the terms with similar $\gamma_{\vec{i}, \vec{j}}$ together, \cref{eqn:OEl_exp} becomes
\begin{align}\label{eqn:pauli_expect_decay}
    \expval{O_{\abs{\mathbb{L}} = l}} = \sum_{k = 1}^{K} A_k \left(1-\gamma_k\right)^l
\end{align}
where $A_k$ is the sum of $2\Re{\alpha_{\vec{i}}^* \alpha_{\vec{j}} \Tr(\rho G_{\vec{i}}^\dagger OG_{\vec{j}})}$ for some subset of $\vec{i}$ and $\vec{j}$. Note that $A_k$ are independent of $l$ and we have $\sum_{k = 1}^K A_k = \expval{O}$.

So far we have only been considering group error channels. However, as shown in \cref{sec:pauli_expec_decay}, by approximating general Pauli channels as the composition of pure group channels, we can prove that decay of the expectation value under general Pauli noise can also be approximated by a sum of exponentials like in \cref{eqn:pauli_expect_decay}.

\cref{eqn:pauli_expect_decay} can be translated into a multi-exponential decay of $\expval{O_\mu}$ over the mean circuit error count $\mu$ using \cref{eqn:err_obs_expand}:
\begin{align}
    \expval{O_\mu} & = e^{-\mu} \sum_{l = 0}^{\infty} \frac{\mu^l}{l!} \sum_{k = 1}^K A_k (1 - \gamma_k)^l\label{eqn:pauli_expect_decomp}\\
    &= \sum_{k = 1}^{K} A_k e^{-  \gamma_k \mu}\label{eqn:comb_extra_poisson}.
\end{align}
This can be rewritten as
\begin{align*}
    \expval{O_\mu} & = \sum_{k = 1}^{K} A_k \left(e^{- \gamma_k}\right)^{\mu} \approx  \sum_{k = 1}^{K} A_k \left( 1- \gamma_k\right)^{\mu}.
\end{align*}
Comparing with \cref{eqn:pauli_expect_decay}, we see that the shape of $\expval{O_\mu}$ and $\expval{O_{\abs{\mathbb{L}} = l}}$ are the same up to the leading order of $\gamma_k$ with the mean circuit error count $\mu$ in place of the circuit error count $l$. 

The simplest shape that we can fit over $\expval{O_\mu}$ is a single exponential decay curve ($K=1$), which is the one we used in exponential extrapolation. We see here that a natural extension of this will be probing $\expval{O_\mu}$ at more than two different error rates and trying to fit them using a sum of exponentials ($K > 1$). The estimate of the error-free observable $\expval{O}$ can then be obtained by substituting $\mu = 0$ into our fitted curve.

From \cref{eqn:comb_extra_poisson}, we see that the $k$th exponential component can only survive up to the mean circuit error count $\mu \sim \frac{1}{\gamma_k}$, thus we can only obtain information about this component by probing at the mean circuit error count $\mu \lesssim \frac{1}{\gamma_k}$. Since $0 \leq \gamma_k \leq 1$, we have $\frac{1}{\gamma_k} \geq 1$ for all $k$, i.e. we should be able to retrieve all exponential components if we can probe enough points within $\mu \lesssim 1$. In practice, there is a minimum mean circuit error count that we can achieve, which we denote as $\mu^*$. To have an accurate multi-exponential fitting, it is essential that for all components with non-negligible $A_k$, we have $\mu^* \lesssim \frac{1}{\gamma_k}$, i.e. none of the critical exponential components has died away at the minimal error rate that we can probe.

\section{Numerical Simulation for Multi-exponential Extrapolation}\label{sec:expec_decay_sim}
In this section, we will try to apply multi-exponential extrapolation to the Fermi-Hubbard model simulation circuit as outlined in Ref.~\cite{caiResourceEstimationQuantum2020}, which consists of local two-qubit components that correspond to different interaction terms and the fermionic swaps. It can be used for both eigenstate preparation and time evolution simulation. We will assume there are $M$ of these two-qubit components, and they all suffer from two-qubit depolarising noise of error probability $p$,  which is just a group channel of the two-qubit Pauli group (without the phase factors). Using \cref{eqn:p_e}, we have 
\begin{align} \label{eqn:pe_Fermi}
    p_\epsilon  = \frac{\abs{\mathbb{E}} - 1}{\abs{\mathbb{E}}} p = \frac{15}{16} p
    & &\Rightarrow& &\mu_{\epsilon} = \frac{15}{16} \mu.
\end{align}

Since we usually know the number of fermions in the system, we can try to verify the fermion number parity symmetry of the output state, which is simply:
\begin{align*}
    S_{\sigma} = \prod_i Z_i
\end{align*}
in the Jordan-Wigner qubit encoding. All the local two-qubit components in the circuit conserve the symmetry $S_{\sigma}$. Hence when we start in a state with the right fermion number, the output state should also have the correct fermion number, enabling us to perform symmetry verification. 

By checking $S_{\sigma} = \prod_i Z_i$, we can detect all error components with \emph{one} $X$ or $Y$ in the local two-qubit depolarising channels since they anti-commute with $S_{\sigma}$. We will remove the other error components in the local two-qubit depolarising channels using quasi-probability. The removed components are those can be generated from the set $\widetilde{\mathbb{Q}} = \{Z_1, Z_2, X_1X_2\}$ following \cref{sec:group_channel}. Thus we have $\abs{\mathbb{Q}} = 2^3 = 8$ and using \cref{eqn:prob_detected} we also have
\begin{align}\label{eqn:pd_Fermi}
    p_{d} = \frac{\abs{\mathbb{E}} - \abs{\mathbb{Q}}}{\abs{\mathbb{E}}} p = \frac{p}{2} & &\Rightarrow& &\mu_{d} = \frac{\mu}{2}.
\end{align}
The resultant noise channel after the application of quasi-probability is given by \cref{eqn:quasi_final_arbit_channel}, which is just a uniform distribution of the two-qubit Pauli errors that are detectable by $S_{\sigma}$. We will call it the \emph{detectable noise}. 

In this section and later in \cref{sec:err_exp_hyp}, we will perform numerical simulations using the circuit for the $2 \times 2$  half-filled Fermi-Hubbard model, which consists of $8$ qubits and $144$ two-qubit gates. The two-qubit gates in the circuit that correspond to interaction terms are parametrised gates with the parameters indicating the strength of the interaction. In our simulation, we will obtain the results for a set of randomly chosen gate parameters (with additional results for another set of random parameters listed in \cref{sec:numeric_2nd}). We will also look at two different error scenarios: depolarising errors and detectable errors. One of them is a group channel while the other is a more general Pauli channel. The measurements that we perform will be the Pauli components of the Hamiltonian, from which we can reconstruct the energy of the output state. The simulations are performed using the Mathematica interface~\cite{jonesQuESTlinkMathematicaEmbiggened2020} of the high-performance quantum computation simulation package QuEST~\cite{jonesQuESTHighPerformance2019}.

\begin{figure}[htbp!]
    \centering
     \centering
    \subfloat[]{\includegraphics[width = 0.45\textwidth]{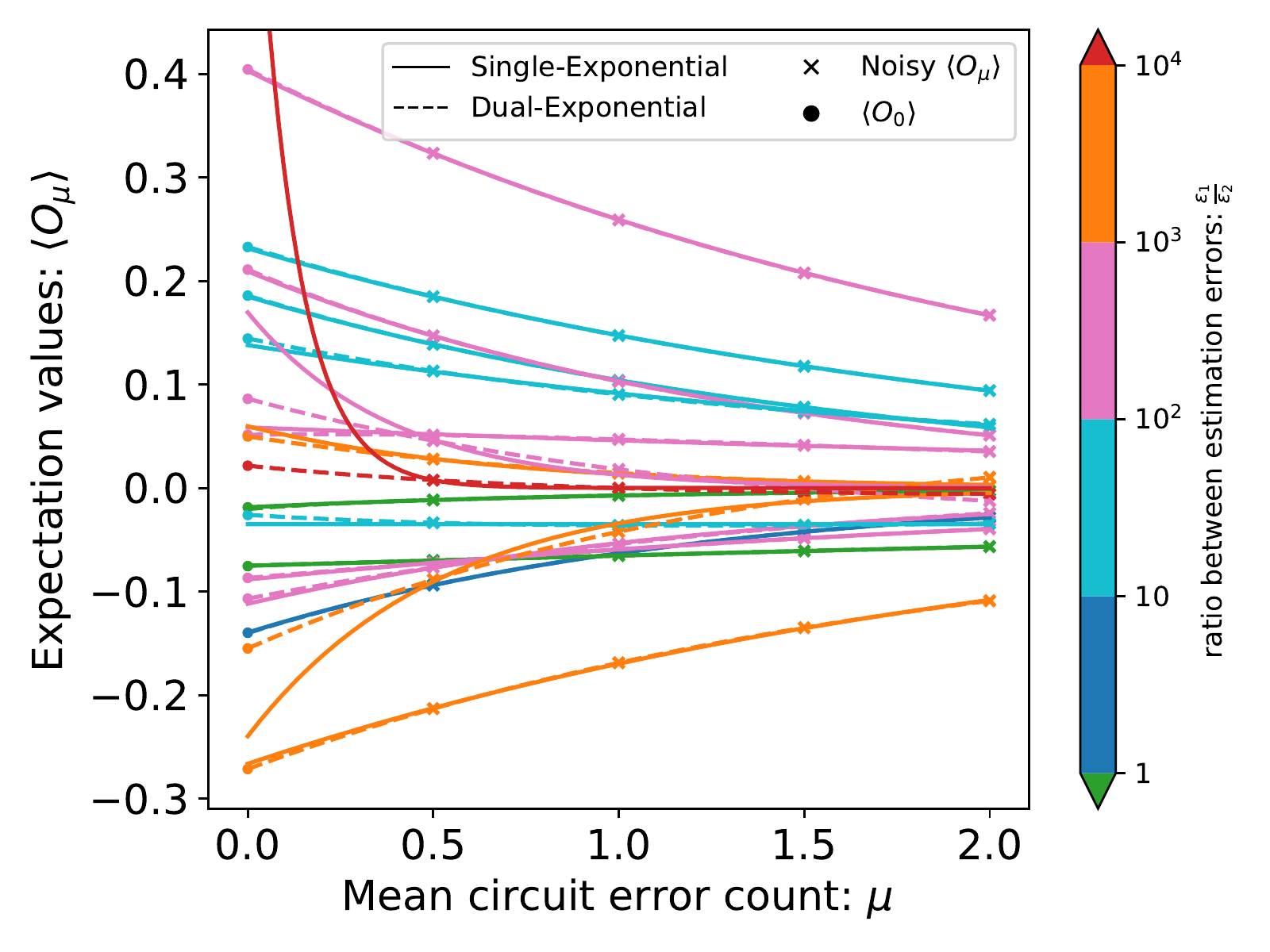}}
    \\
    \subfloat[]{\includegraphics[width = 0.45\textwidth]{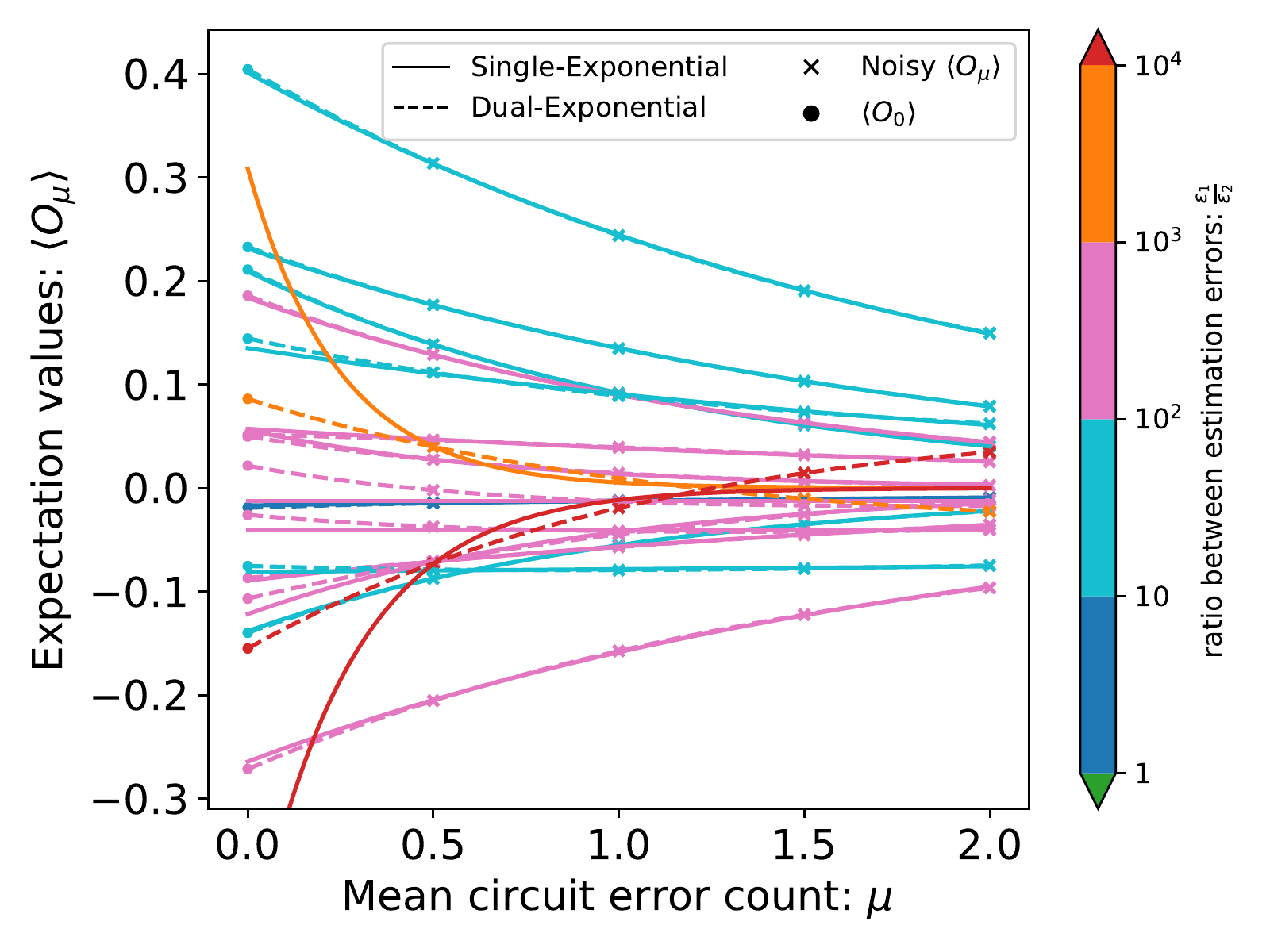}}
    \caption{\textbf{Comparison between single-exponential extrapolation and dual-exponential extrapolation in a $8$-qubit simulation.} Plots showing the noisy expectation values of different Pauli observables under (a) depolarising noise and (b) detectable noise obtained at the four mean circuit error counts $\mu = 0.5,\ 1,\ 1.5 ,\ 2$ (cross markers). The single- and dual-exponential extrapolation curves fitted to the data points are represented by the solid and dashed lines, respectively. The circular markers lie at $\mu = 0$ and denote the true noiseless expectation values. Different colours represent different ratios between the estimation bias of using single- and dual-exponential extrapolation (e.g. orange means that for the given observable, the estimation bias of single-exponential extrapolation is between $10^3$ to $10^4$ times larger than that of dual-exponential extrapolation).}
    \label{fig:one_exp_vs_two_exp}
\end{figure}

In \cref{fig:one_exp_vs_two_exp}, we have plotted the noisy expectation values $\expval{O_\mu}$ for each Pauli observable at the mean circuit error counts $\mu = 0.5,\ 1,\ 1.5 ,\ 2$. When we perform multi-exponential extrapolation on them, we find that all of the observables can be fitted using a sum of at most two exponentials, even though there should be very few symmetries in our circuits since they are generated from a set of random parameters. We now can proceed to compare the absolute bias in the estimate of the noiseless expectation values ($\mu = 0$) using dual-exponential extrapolation against that using the conventional single-exponential extrapolation for all of the noisy observables in \cref{fig:one_exp_vs_two_exp}. The absolute biases of single- and dual-exponential extrapolation are denoted as $\epsilon_{1}$ and $\epsilon_{2}$, respectively, and different colours in the plots correspond to different estimation bias ratios $\nicefrac{\epsilon_{1}}{\epsilon_{2}}$.

\begin{figure}[htbp!]
    \centering
    \includegraphics[width = 0.45\textwidth]{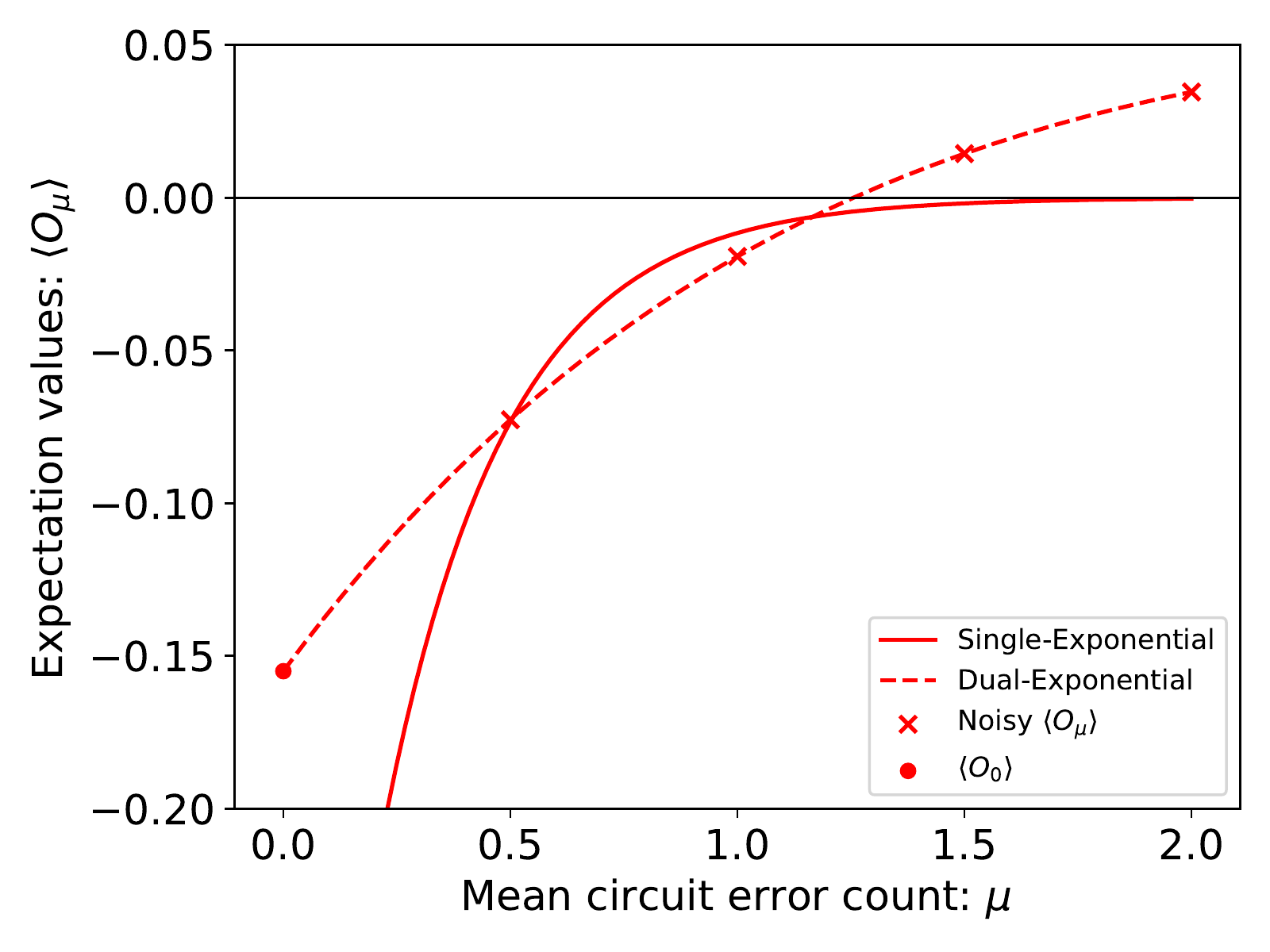}
    \caption{\textbf{A noisy Pauli observable from \cref{fig:one_exp_vs_two_exp} (b) that cannot be fitted well using single exponential decay since it crosses over the x-axis.} Here we have plotted the noisy expectation values at the four mean circuit error counts $\mu = 0.5,\ 1,\ 1.5 ,\ 2$ (cross markers). The single- and dual-exponential extrapolation curves fitted to the data points are represented by the solid and dashed lines, respectively. The circular markers lie at $\mu = 0$ and denote the true noiseless expectation values.}
    \label{fig:why_two_exp}
\end{figure}

For 32 out of the 34 observables we plotted, dual-exponential extrapolation can achieve a smaller estimation bias than single-exponential extrapolation ($ \nicefrac{\epsilon_{1}}{\epsilon_{2}} > 1$). Within the two cases that dual-exponential extrapolation is outperformed (the green curves in \cref{fig:one_exp_vs_two_exp}), one of them is the case in which single- and dual-exponential extrapolation both achieve very small estimation bias of similar order. The other remaining case with larger $\epsilon_{2}$ relative to $\epsilon_{1}$ is mainly due to the small magnitude of its true expectation value, which lead to large uncertainties in the fitting parameters. It is also because we are only using the bare minimum of 4 data points to fit a dual-exponential curve with 4 free parameters and thus the problem may be alleviated by simply probing at more error rates to obtain more data points. On the other hand, there are also a few cases in which the $\epsilon_{1}$ are exceptionally large (e.g. certain orange and red curves in \cref{fig:one_exp_vs_two_exp} (a) and (b)). These are usually observables whose decay curves have extrema and/or crossing over the x-axis, thus it is \emph{impossible} to get a good fit with a single-exponential curve. We have zoomed into one such observable in \cref{fig:why_two_exp}. For these observables, dual-exponential extrapolation can still perform extremely well and achieve $\epsilon_{2} \sim 10^{-5}$, which is up to tens of thousands times lower than $\epsilon_{1}$: $ \nicefrac{\epsilon_{1}}{\epsilon_{2}} \sim 10^4$.

\begin{table}[htbp!]
   \centering
   \begingroup
   \renewcommand{\arraystretch}{1.5}
   \begin{tabular}{lcc}\toprule
       $\overline{\epsilon}_{1}, \overline{\epsilon}_{2} / 10^{-4}$ &Depolarising  &Detectable \\ \hline
       Single-exp& 150  &74  \\
       Dual-exp&  1.0   & 1.0\\
       \botrule
   \end{tabular}
   \endgroup
    \caption{The bias in the single- and dual-extrapolation estimates averaged over observables within each plots in \cref{fig:one_exp_vs_two_exp} excluding observables with exceptionally large $\epsilon_{1}$ or $\epsilon_{2}$. The entries are in the unit of $10^{-4}$. }
    \label{tab:two_exp_err}
\end{table}

Now we will exclude the few observables above with exceptionally large $\epsilon_{1}$ or $\epsilon_{2}$ and take the average of the remaining $\epsilon_{1}$ and $\epsilon_{2}$ to obtain a more representative performance of dual-exponential extrapolation against single-exponential extrapolation. This is shown in \cref{tab:two_exp_err}, from which we see that by using dual-exponential extrapolation instead of single-exponential extrapolation, we can achieve tens or even a hundred times reduction in estimation bias across both noise models. Note that in \cref{fig:one_exp_vs_two_exp} it appears to the eye that the true (noiseless) expectation values, marked by filled circles, never deviate from the dual-exponential (dashed) lines. In fact there are minute discrepancies as specified in \cref{tab:two_exp_err}, but the extrapolation is remarkably successful.

\section{Combination of Error Mitigation Techniques}\label{sec:comb_err_miti}
We have shown that extrapolating using multi-exponential curve can be very effective assuming Pauli noise. Besides the shape of the extrapolation curve, the other key component to error extrapolation is the way to tune the noise strength. Previously, noise are \emph{boosted} by increasing gate pulse duration~\cite{temmeErrorMitigationShortDepth2017}, applying additional gates that cancel each other~\cite{giurgica-tironDigitalZeroNoise2020a} or simulating the noise using random gate insertion~\cite{liEfficientVariationalQuantum2017}. There are various practical challenges associated with these noise-boosting techniques, and furthermore as discussed at the end of \cref{sec:err_extrapolation}, data points at boosted error rates may not contain enough information for effective extrapolation. Now as shown in \cref{sec:qua_prob}, we can actually use quasi-probability to reduce the error rate and obtain a set of data points with \emph{reduced} noise strength, which can then be used for extrapolation. We should expect a smaller estimation bias by using data points with reduced noise strength rather than boosted noise strength, but the sampling cost will also increase due to the use of quasi-probability. For the special case of two-point extrapolation using a single-exponential curve with the two data points at the unmitigated error rate $\mu$ and the quasi-probability-suppressed error rate $\nu$, the total sampling cost as shown in \cref{sec:QE_cost} is
\begin{align}\label{eqn:cost_qua_extrapolate}
    C_{QE} \sim 2\frac{\lambda^2 e^{\frac{2}{\lambda} \left[\gamma \mu + 2\left(\lambda - 1\right) \mu_\epsilon\right]}  +  e^{2 \gamma \mu}}{\left(\lambda - 1\right)^2}
\end{align}
with $\mu = \lambda \nu$. We will call this special case \emph{quasi-probability with exponential extrapolation}.

As discussed in \cref{sec:err_extrapolation}, the number of exponential components in the multi-exponential extrapolation can be reduced with increased degree of symmetry in the circuit and/or if the error channels are group channels. Hence, besides using quasi-probability for error suppression, we can also use quasi-probability to transform the error channels into group errors and/or errors of similar form for easier curve fitting in the extrapolation process. 

Moving on, we may wish to combine symmetry verification and quasi-probability.  We can first apply quasi-probability to transform all the error channels in the circuit with the total mean error count $\mu$ into \emph{detectable} error channels with a total mean error count $\mu_d$. After that, we can apply symmetry verification, but note that the additional quasi-probability operations may contain gates that take us from one symmetry space to another, for which we need to adjust our symmetry verification criterion accordingly.  As discussed in \cref{sec:nisq_limit}, an even number of occurrence of local detectable errors can still escape the symmetry test and lead to circuit errors. We can further suppress them by applying additional quasi-probability operations as shown in \cref{sec:comb_ver_qua}. Alternatively, we can also try to remove these remaining errors by applying error extrapolation as we will see below.

After using quasi-probability to transform all local errors into local detectable errors with the mean circuit error count $\mu_d$, performing symmetry verification will split the circuit runs into two partitions, one has an even number of detectable errors occurring and will pass the symmetry test, the other has an odd number of detectable errors occurring and will fail the symmetry test.
Consequently, the noisy observable expectation value (\cref{eqn:pauli_expect_decomp}) can also be split into the weighted sum of these two partitions:
\begin{align}\label{eqn:exp_fraction}
    \expval{O_{\mu_d}} = e^{-\mu_d}\left(\cosh(\mu_d)\expval{O_{c, \mu_d}} + \sinh(\mu_d)\expval{O_{s, \mu_d}}\right)
\end{align}
in which $e^{-\mu_d}\cosh(\mu_d)$ and $e^{-\mu_d}\sinh(\mu_d)$ are the probability to have even and odd number of errors occurring in the circuit, respectively. And $\expval{O_{c, \mu_d}}$, $\expval{O_{s, \mu_d}}$ are the corresponding expectation values in these cases with
\begin{equation}\label{eqn:cosh_sinh_K}
    \begin{aligned}
        \expval{O_{c, \mu_d}} = \frac{1}{\cosh(\mu_d)}\sum_{k = 1}^K A_k \cosh( \left(1- \gamma_k\right) \mu_d),\\
        \expval{O_{s, \mu_d}} = \frac{1}{\sinh(\mu_d)}\sum_{k = 1}^K A_k \sinh(\left(1- \gamma_k\right) \mu_d).
    \end{aligned}
\end{equation}
We will consider the case that the decay of our expectation value $\expval{O_{\mu_d}}$ over increased mean circuit error count $\mu_d$ follows a single exponential curve ($K = 1$) for simplicity, we then have:
\begin{equation}\label{eqn:cosh_sinh}
    \begin{aligned}
        \expval{O_{c, \mu_d}} = \expval{O} \frac{\cosh(\left(1- \gamma\right) \mu_d)}{\cosh(\mu_d)}\\
        \expval{O_{s, \mu_d}} = \expval{O} \frac{\sinh(\left(1- \gamma\right) \mu_d)}{\sinh(\mu_d)}
    \end{aligned}
\end{equation}
which gives:
\begin{equation}
    \begin{aligned}\label{eqn:get_O0}
        \expval{O} &= \text{sgn}\left(\expval{O_{c, \mu_d}}\right) \\
        & \quad \times \sqrt{\expval{O_{c, \mu_d}}^2\cosh[2](\mu_d) - \expval{O_{s, \mu_d}}^2\sinh[2](\mu_d)}.
    \end{aligned}
\end{equation}
Note that we have used the fact that $\expval{O_{c, \mu_d}}$ and $\expval{O}$ have the same sign since $ 1- \gamma > 0$ and $\mu_d >0$. Here with the help of symmetry verification and quasi-probability, we can now obtain an estimate of the error-free expectation value $\expval{O}$ by combining the expectation value of the passed runs and failed runs at one error rate $\mu_d$ instead of combining the expectation value of runs at different error rates in the conventional error extrapolation. Note that here we have assumed that we know the value of the mean detectable circuit error count $\mu_d$, which needs be known before we can apply the quasi-probability step anyway. The method we employed in \cref{eqn:get_O0} will be called \emph{hyperbolic extrapolation}.

As derived in \cref{sec:hyper_cost}, the sampling cost factor of hyperbolic extrapolation is
\begin{align}\label{eqn:sym_extrapolate_cost}
    C_{H, \mu_d} & = \cosh(2\left(1- \gamma\right) \mu_d)\cosh(\mu_d)e^{\mu_d}.
\end{align}
To combine all three error mitigation techniques, we first use quasi-probability to remove the error components that are undetectable by symmetry verifications. Applying symmetry verification will then split the circuit runs into two sets: runs with even number of errors and runs with odd number of errors, obtaining two separate expectation values. Using our understanding about the decay of the expectation value from our study of error extrapolation, we can simply combine these two erroneous expectation values and obtain the error-free expectation value. The full process is called \emph{quasi-probability with hyperbolic extrapolation}, and the corresponding total sampling cost factor can be obtained using \cref{eqn:partial_qua_cost} and \cref{eqn:sym_extrapolate_cost}:
\begin{align}\label{eqn:comb_opt_cost}
    C_{QH}(\gamma) = C_{Q,\mu_d} C_{H, \mu_d} = e^{4\mu_\epsilon} \frac{\cosh(\mu_d)\cosh(2\left(1- \gamma\right) \mu_d)}{e^{3\mu_d}}.
\end{align}
We note that this is always smaller than the cost of pure quasi-probability $C_{Q,0} = e^{4\mu_\epsilon}$. Its cost saving over pure quasi-probability will increase with the increase of $\gamma$.

\section{Numerical Simulation for Combination of Error Mitigation Techniques}\label{sec:err_exp_hyp}

\begin{figure}[htbp]
    \centering
    \subfloat[]{\includegraphics[width = 0.45\textwidth]{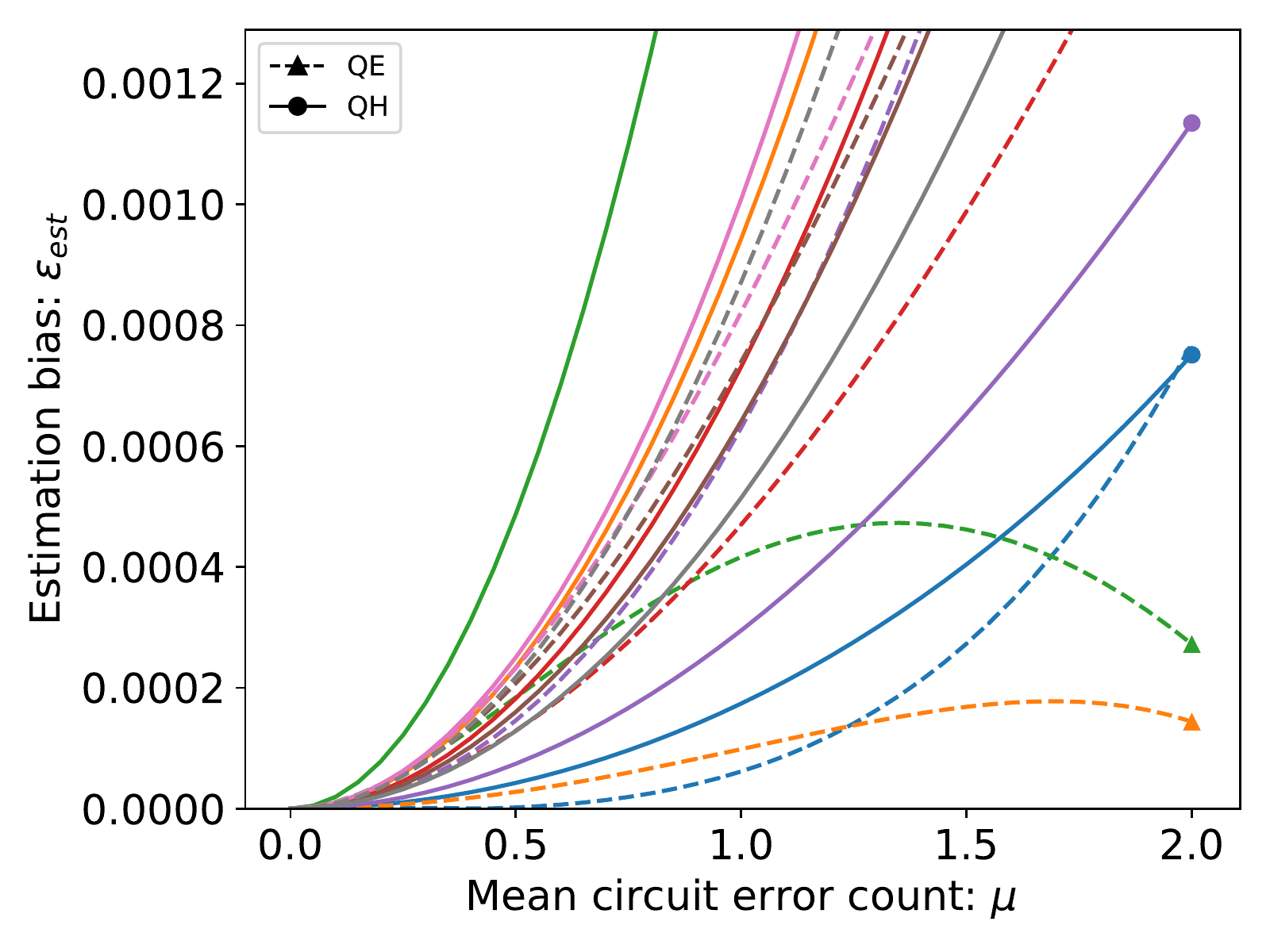}}
    \\
    \subfloat[]{\includegraphics[width = 0.45\textwidth]{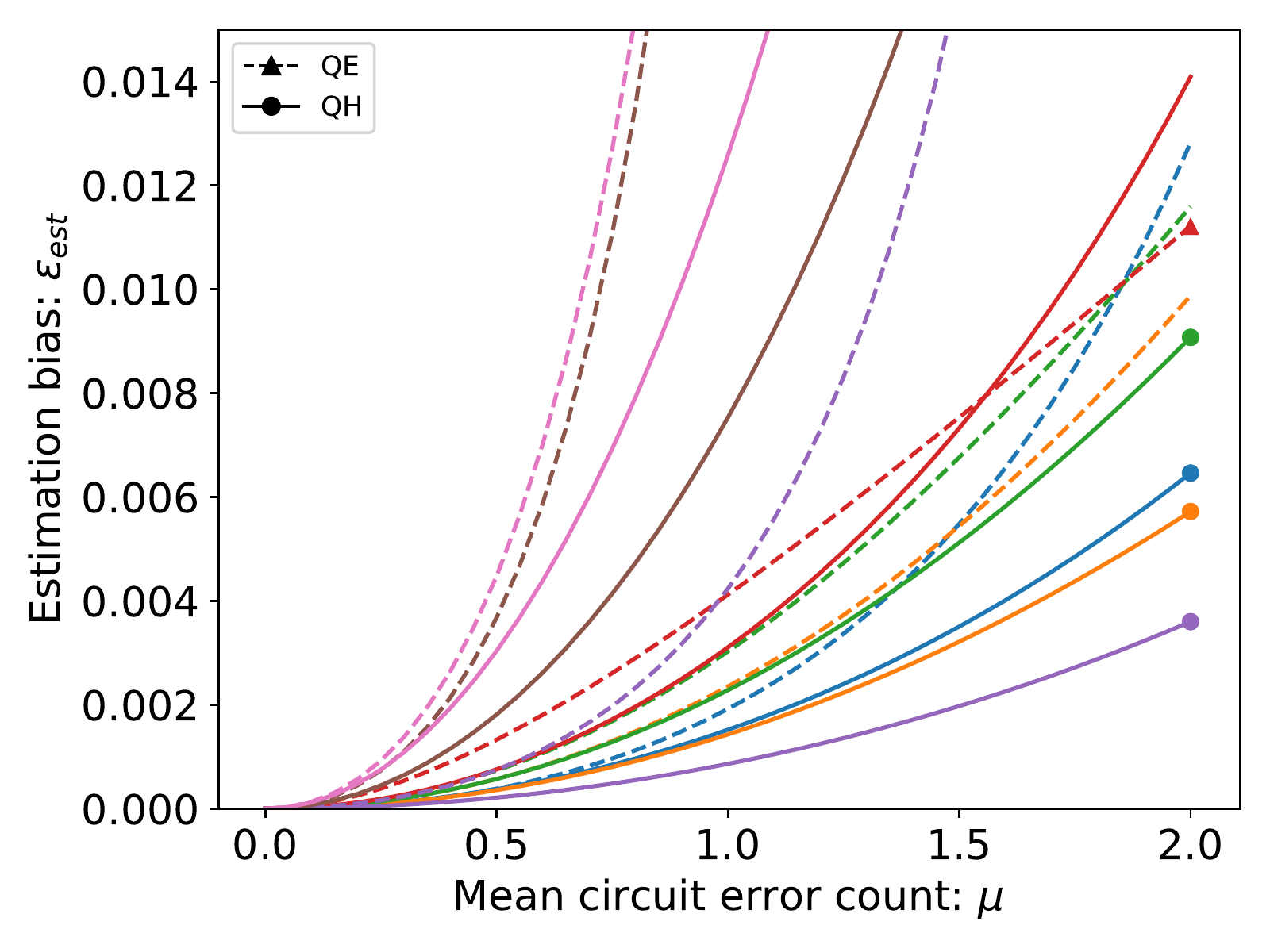}}
    \caption{\textbf{Comparison of the biases in the error-mitigated expectation values between quasi-probability with exponential extrapolation (QE) and quasi-probability with hyperbolic extrapolation (QH) in a 8-qubit simulation.} Plots showing (a) observables following single-exponential decay and (b) observables following dual-exponential decay. Within each plot, different colours represent different observables. The solid lines denote QH, while the dashed lines denote QE. At the mean circuit error counts $\mu = 2$, for each observable, we use markers to denote the method that has lower estimation bias out of the two. For a given observable, circular markers denote lower estimation bias when using QH, while triangle markers denote lower estimation bias when using QE.}
    \label{fig:est_err}
\end{figure}

In this section, we will compare the performance of quasi-probability with exponential extrapolation (QE) and quasi-probability with hyperbolic extrapolation (QH) discussed in \cref{sec:comb_err_miti}. Similar to \cref{sec:expec_decay_sim}, we will perform Fermi-Hubbard model simulation with local two-qubit depolarising noise with a mean circuit error count $\mu$. The symmetry we will used in QH is the fermionic number parity symmetry, which means that the resultant mean detectable circuit error count after we apply the quasi-probability step in QH will be $\mu_d = \frac{\mu}{2}$ following \cref{eqn:pd_Fermi}. For the quasi-probability in QE in this section, we will keep it at the same strength as that in QH, which means that they have the same resultant circuit error rate: $\nu = \frac{\mu}{2}$. Note that even though resultant channels after the partial quasi-probability in both QE and QH give the same mean circuit error count $\nu = \frac{\mu}{2} = \mu_d$, in one case the resultant noise is still depolarising while in the other case the resultant noise is locally detectable. In this section, we will assume the quasi-probability process is performed perfectly. Recall that for simplicity we have only explicitly derived QH under the assumption that the observable follows a single-exponential decay, so for a fair comparison the QE method in this section will also only employ single-exponential extrapolation. However as we will see later, even when we look at observables that follow a dual-exponential decay, which breaks our assumptions above, QH can still achieve robust performance. 

As shown in \cref{fig:one_exp_vs_two_exp}, for our example circuits, some observables can be fitted well enough using single-exponential decay curves while the other observables can only be fitted well using dual-exponential decay curves. We will call these two types of observables \emph{single-exponential observables} and \emph{dual-exponential observables}, respectively. In \cref{fig:est_err}, we have plotted the absolute estimation bias $\epsilon_{est}$ using the two different extrapolation techniques for the single-exponential and dual-exponential observables. First, we can see that the estimation biases for the dual-exponential observables are almost one order of magnitude higher than the biases for the single-exponential observables. This should not come as a surprise since both single-exponential extrapolation and hyperbolic extrapolation are derived under the assumptions of single-exponential observables. At the mean circuit error counts $\mu = 2$, for each observable, we have used markers to label the method that can achieve a lower estimation bias out of the two. We see that the number of single-exponential observables that can achieve a lower estimation bias using QE is comparable to that of QH. On the other hand, almost all dual-exponential observables can achieve a lower estimation bias using QH.

\begin{table}[htbp!]
    \centering
    \begingroup
    \renewcommand{\arraystretch}{1.5}
    \subfloat[]{
        \begin{tabular}{cccc}\toprule
            $\overline{\epsilon}_{est}/ 10^{-4}$& \  1-Exp. Obs. \  & \  2-Exp. Obs. \ & \  All Obs. \ \\ \hline
            QE &5.1&100&53\\
            QH &7.8&56&32\\
            \botrule
        \end{tabular}
    }\\
    \subfloat[]{
        \begin{tabular}{cccc}\toprule
            $\overline{\epsilon}_{est}/ 10^{-3}$& \  1-Exp. Obs. \  & \  2-Exp. Obs. \ & \  All Obs. \ \\ \hline
            QE &1.8&82&39\\
            QH &3.2&20&11\\
            \botrule
        \end{tabular}
    }
    
    \endgroup
    \caption{The biases in the error-mitigated estimates using QE and QH averaged over single-exponential, dual-exponential and all observables at the mean circuit error counts (a) $\mu = 1$ and (b) $\mu = 2$. The entries in (a) and (b) are in the units of $10^{-4}$ and $10^{-3}$, respectively.}
    \label{tab:est_err}
\end{table}

In \cref{tab:est_err}, we further calculate the average estimation bias of single-exponential, dual-exponential and all observables separately at the circuit error rate $\mu = 1, 2$, which re-confirm all of our observations above.
We see that the estimation bias of QE is lower than that of QH for single-exponential observables, and on the other hand QH can achieve a lower estimation bias for dual-exponential observables. In other words, the performance of QH is more robust against whether the observable is single-exponential or not. When looking at the estimation bias averaged over all observables, we see the estimation bias of QH is always lower than QE and can be $4$ times smaller than QE at $\mu = 2$. The all-observable averages can be more indicative about the practical performance of the mitigation techniques since in experiments we do not know whether a given observable should be fitted with single-exponential or not beforehand. 

There is another added layer of robustness when we try to apply QH instead of QE to multi-exponential observables when we look back at the hyperbolic extrapolation equation \cref{eqn:get_O0}. We can see that if the shape of the observable is far off from a single-exponential decay, then this might lead to a negative number in the square root of \cref{eqn:get_O0}, allowing us to realise that we need to probe at more error rates to perform multi-exponential extrapolation instead, and avoiding performing a bad extrapolation with a very large bias. In the simulation, we indeed identify a few observables that we cannot perform QH on. For these observables, we can still perform QE, but it will lead to huge biases in the estimates. These observables have been excluded in our comparison between QE and QH.

\begin{figure*}[htbp]
    \centering
    \subfloat[$\gamma = 1$.]{\includegraphics[width = 0.45\textwidth]{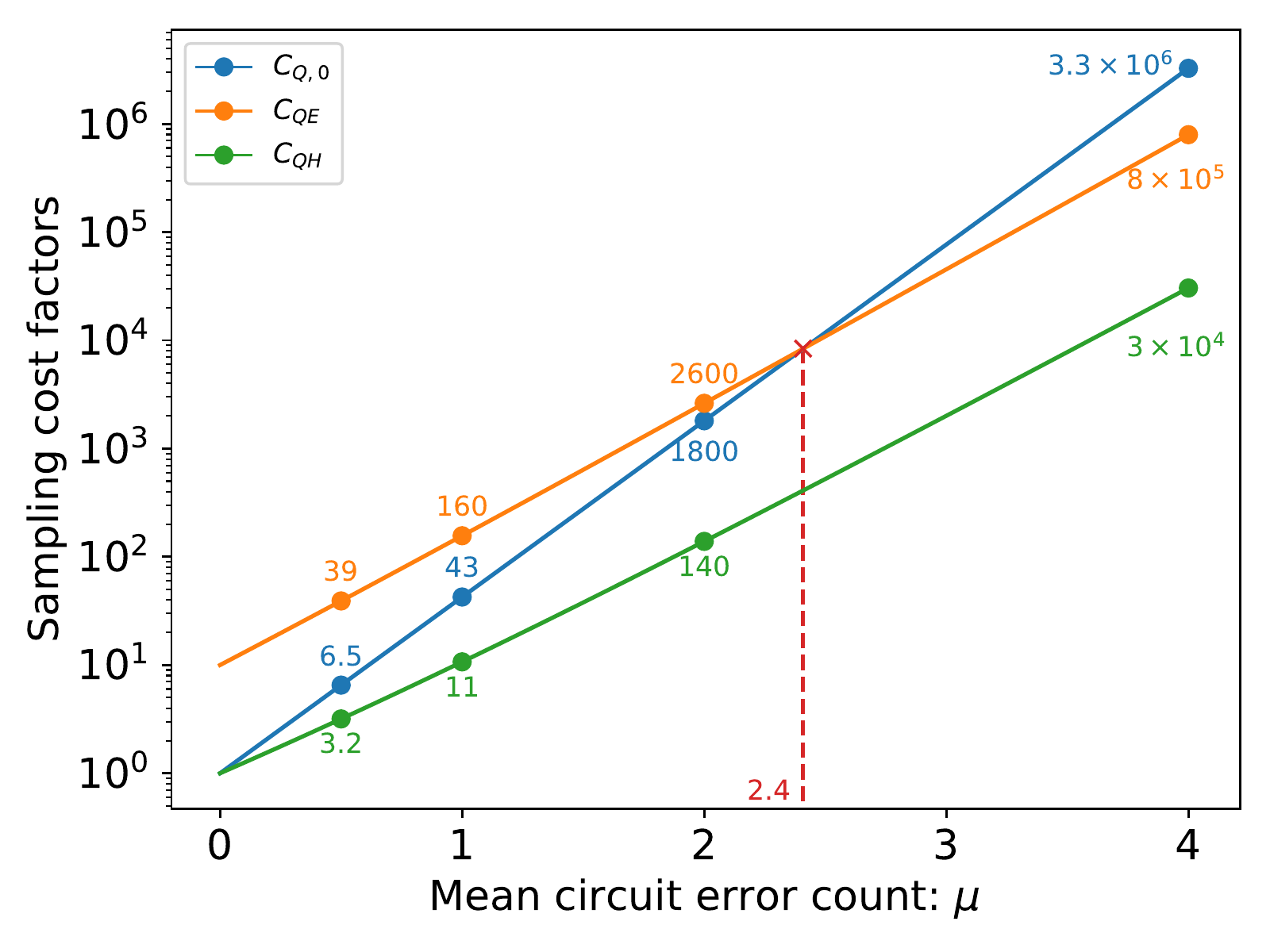}}
    \quad
    \subfloat[$\gamma = 0$.]{\includegraphics[width = 0.45\textwidth]{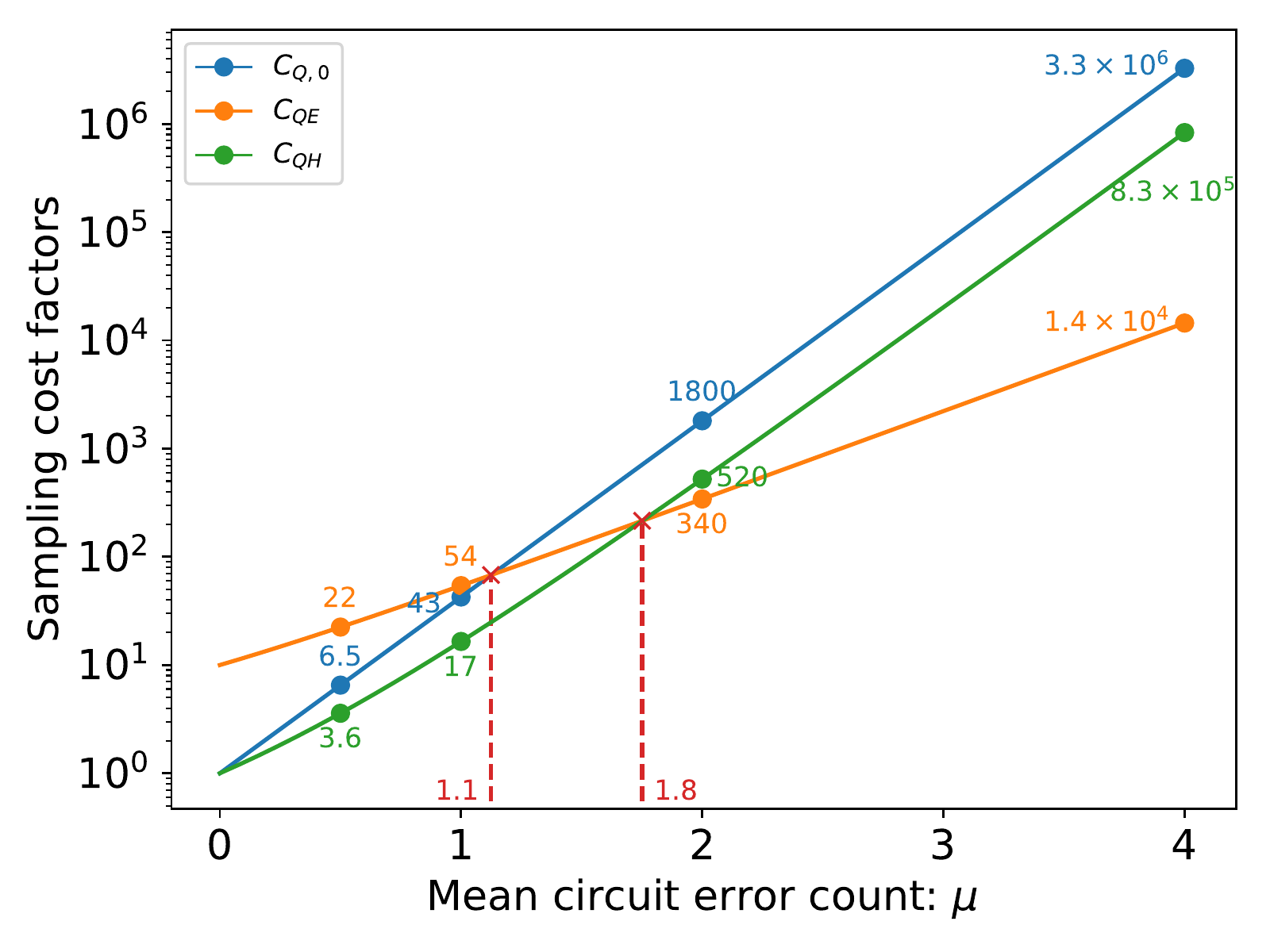}}\\
    \subfloat[$\gamma = 0.5$.]{\includegraphics[width = 0.6\textwidth]{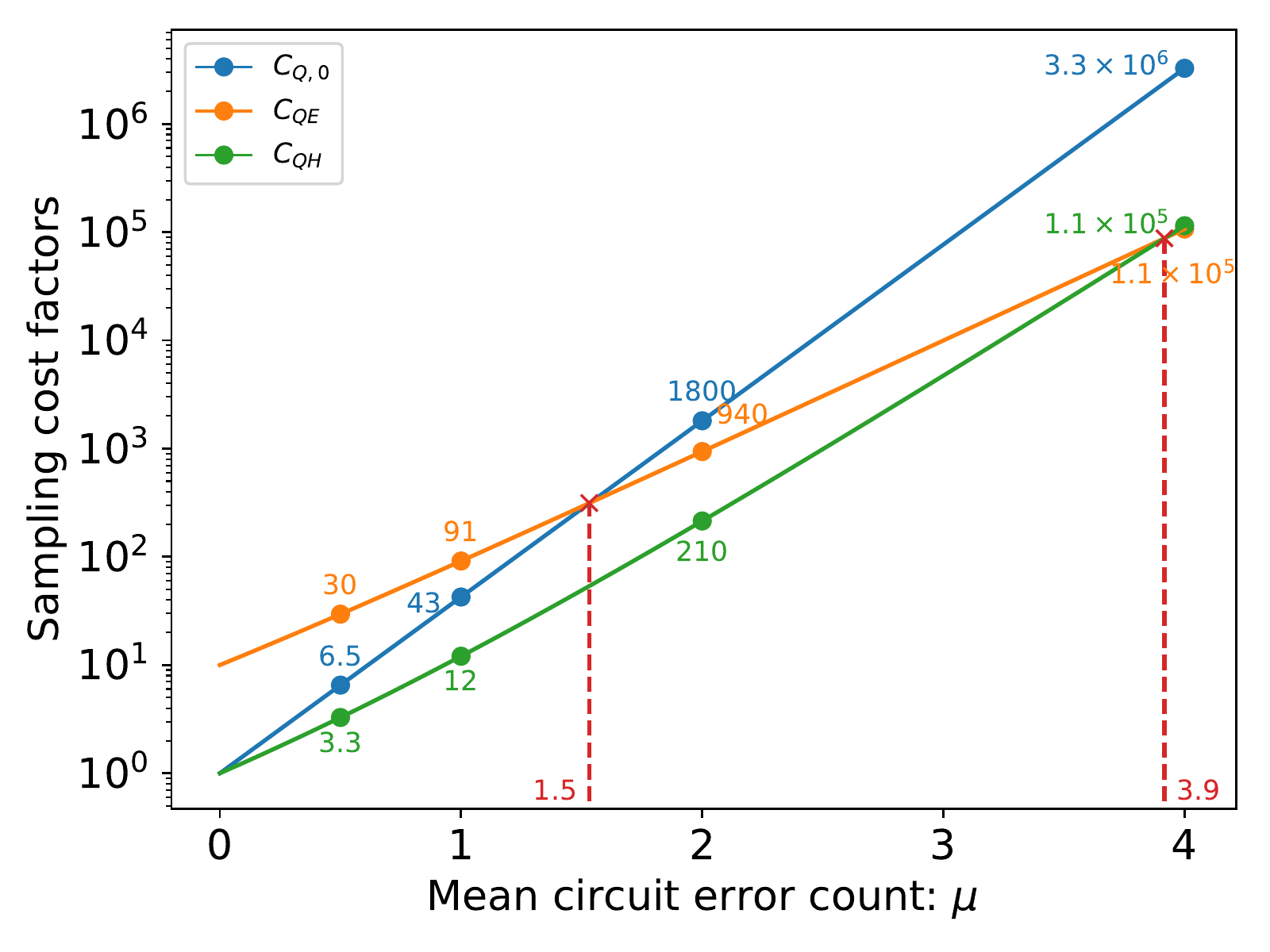}}
    \caption{
        \textbf{Comparison of the sampling cost factors between pure quasi-probability (Q), quasi-probability with exponential extrapolation (QE) and quasi-probability with hyperbolic extrapolation (QH).} Plots showing different noisy observable decay rate (a) $\gamma = 1$, (b) $\gamma = 0$ and (c) $\gamma = 0.5$. We have labelled the values of the lines at the mean circuit error counts $ \mu = 0.5,\ 1,\ 2,\ 4$. We have also labelled the intersects between lines of different methods using red markers.
    }
    \label{fig:extrapolate_cost}
\end{figure*}

Using \cref{eqn:comb_opt_cost}, \cref{eqn:pe_Fermi} and \cref{eqn:pd_Fermi}, the sampling cost factor of performing QH in our example circuit is
\begin{align*}
    C_{QH}(\gamma) &= e^{\frac{9}{4}\mu} \cosh(\frac{\mu}{2}) \cosh(\left(1- \gamma\right) \mu),
\end{align*}
where $\gamma$ is the decay rate of the observable expectation values under noise.

Using \cref{eqn:cost_qua_extrapolate}, \cref{eqn:pe_Fermi} and $\nu = \mu_d = \frac{\mu}{2} \Rightarrow \lambda = 2$, the sampling cost factor of performing QE is:
\begin{align*}
    C_{QE} &= 2\left(4 e^{\left(\gamma + \frac{15}{8}\right) \mu}  +  e^{2 \gamma \mu}\right).
\end{align*}

For comparison purpose, we also write down the sampling cost factor for removing all the errors using quasi-probability given by \cref{eqn:pure_qua_cost}:
\begin{align*}
    C_{Q, 0} \approx e^{4\mu_{\epsilon}} = e^{\frac{15}{4}\mu}.
\end{align*}

The comparison between $C_{Q, 0}$, $C_{QE}(\gamma)$ and $C_{QH}(\gamma)$ at different $\gamma$ is plotted in \cref{fig:extrapolate_cost}. We can see that $C_{QH}(\gamma)$ is always lower than $C_{Q, 0}$ across all $\mu$ and $\gamma$, i.e. we can always get a sampling cost saving by applying QH instead of pure quasi-probability, which is also proven in \cref{sec:comb_err_miti}. On the other hand, at $\gamma = 1$, $C_{QE}$ is larger than $C_{QH}$ for all $\mu$ and larger than $C_{Q, 0}$ for $\mu < 2.4$. As $\gamma$ decreases, $C_{QH}(\gamma)$ will increase while $C_{QE}(\gamma)$ will decrease. Thus they naturally complement each other as QH will be more suitable for large-$\gamma$ error mitigation while QE will be more suitable for small-$\gamma$ error mitigation. At $\gamma = 0$, we see that $C_{QE}$ becomes lower than both $C_{QH}$ and $C_{Q, 0}$ at $\mu > 1.8$.

The average fitted $\gamma$ of all the single-exponential observables within each plot in \cref{fig:one_exp_vs_two_exp} all lie within the range $0.5 - 0.6$. Hence, we will now focus on the $\gamma = 0.5$ plot in \cref{fig:extrapolate_cost} to get an indication of the practical sampling costs of implementing different mitigation techniques.  

At $\mu = 1$, the sampling cost factor of quasi-probability is $43$. QE requires a higher sampling cost, thus there is no point performing QE since pure quasi-probability can remove all the noise perfectly in theory with a lower cost. Compared to quasi-probability, QH can reduce the cost by more than $70 \%$ while still achieving the small estimation bias $\overline{\epsilon}_{QH} \sim 3 \times 10^{-3}$ shown in \cref{tab:est_err}. In order for quasi-probability to have any advantages over QH, we must sample enough times such that the shot noise of pure quasi-probability is smaller than the estimation bias of QH (more rigorous arguments in \cref{sec:comp_miti}), which will require
$N^* \sim \nicefrac{C_{Q, 0}}{\overline{\epsilon}_{QH}^2} \approx 4.3\times 10^6$
samples for each observable. Therefore in practice, QH could be the preferred method over pure quasi-probability as it is challenging to sample more than $N^*$ for each observable within reasonable runtime constraints~\cite{caiResourceEstimationQuantum2020}. 

At $\mu = 2$, now the sampling cost factor of quasi-probability is $1800$, which is hardly practical. QE can reduce this sampling cost by half while achieving estimation bias around $4 \times 10^{-2}$ (\cref{tab:est_err}), and QH can reduce this sampling cost by almost $90\%$ while achieving estimation bias around $1 \times 10^{-2}$ (\cref{tab:est_err}), thus they both would be preferred over pure quasi-probability in practice following similar arguments in the $\mu = 1$ case. We also see that QH outperforms QE in terms of both sampling cost and estimation bias at $\mu = 2$, and thus would be preferred over QE. The cost of QE will only become lower than QH at $\mu = 3.9$, however at this point, neither of their sampling costs are likely to be practical. 

\section{Discussion}\label{sec:conclusion}
In this Article, we have recapped and studied the mechanism and performance of three of the most well-known error mitigation techniques: symmetry verification, quasi-probability and error extrapolation under Pauli noise. By introducing the concepts of group errors and NISQ limits, we managed to prove that the change of the expectation value of a Pauli observable with increased Pauli noise strength can be approximated using multi-exponential decay, enabling us to extend exponential error extrapolation to multi-exponential extrapolation. We then performed 8-qubit numerical simulations for Fermi-Hubbard simulation under two different Pauli noise models, finding that the decay of their Pauli expectation values can all be fitted using single- or dual-exponential curves, confirming our earlier proof of multi-exponential decay. Using the same circuits, we performed dual-exponential extrapolation by probing at four different error rates, which is minimal number of data points required, and managed to obtain a low estimation bias of $\lesssim 10^{-4}$ for almost all $34$ observables except for one fringe case. In our simulations, the estimation bias of dual-exponential extrapolation are on average $\sim 100$ times lower than that of single-exponential extrapolation, with the maximum factor of bias reduction reaching $\sim 10^4$.

We then proceeded to combine different error mitigation techniques in the context of well-characterised local Pauli noise. Instead of using quasi-probability to completely remove all the noise, we can use it to suppress the noise strength and perform error extrapolation, which is named \emph{quasi-probability with exponential extrapolation} (QE). Alternatively, we can use quasi-probability to remove the local undetectable noise and perform symmetry verification. Then instead of discarding all the circuit runs that fail the symmetry test, we have developed a way to recombine the expectation values of the `failed' and `passed' runs to obtain an estimate of the noiseless observable. The full combined method is called \emph{quasi-probability with hyperbolic extrapolation} (QH). Note that both QE and QH are free of the requirement to adjust the hardware error rate despite the name `extrapolation'. By performing $8$-qubit Fermi-Hubbard model simulations under local depolarising noise and using the fermionic number parity symmetry, we found that QH outperforms QE in terms of both estimation bias and sampling costs for almost all cases. When compared to pure quasi-probability, QH can achieve factor-of-4 and factor-of-9 sampling cost savings at the mean circuit error count $\mu = 1$ and $\mu = 2$, respectively, while still maintaining low estimation bias of $10^{-3} \sim 10^{-2}$. Hence, QH would outperform pure quasi-probability in our examples unless we obtain an impractical number of samples (more than millions) per observable.

QH is derived under the assumption that the observables decay along single-exponential curves with increased noise. Our simulation shows that QH can be robust against violation of this assumption when applied to dual-exponential observables. However, such robustness may not persist with a further increase in the number of exponential components. A multi-exponential version of QH can be done through probing at more error rates and fitting \cref{eqn:cosh_sinh_K} to the data. Alternatively, instead of probing at more error rates, we can also try to verify more symmetries. In such a way, we can obtain a set of expectation values corresponding to different verification syndromes for the multi-exponential version of the hyperbolic fitting. An example can be using the separate fermion number parity symmetries for each spin subspace, which will lead to expectation values corresponding to the four possible verification syndromes. However, how to recombine these expectation values in the case of multiple symmetries and how to use quasi-probability to transform the local error channels into the suitable forms for such a recombination is not a simple extension of the single-symmetry case we considered. 

In our derivation, the number of exponential components in the expectation value decay curve in \cref{eqn:pauli_expect_decay} is expected to scale exponentially with the number of gates. However, in our simulations, we fitted at most two exponential components for each of the observable decay curves. More analysis is needed to bridge the gap between the expected and the actual number of exponential components required, possibly based on the symmetry of the circuit. This will help us understand how the number of exponential components scales with the system size, enabling us to gauge the performance and the costs of scaling up the multi-exponential extrapolation method. It might be useful to draw ideas from non-Clifford randomized benchmarking~\cite{crossScalableRandomisedBenchmarking2016, helsenNewClassEfficient2019, flammiaEfficientEstimationPauli2020}, in which multi-exponential decay is also employed for the fitting of the fidelity curves. When applying multi-exponential extrapolation in practice, we might want to develop Bayesian methods to determine whether we need to probe at more error rates, which error rates to probe, and whether to change the number of exponential components of our fitted curve based on the existing data. This has been done in the context of randomized benchmarking~\cite{granadeQInferStatisticalInference2017} and it would be interesting to see its performance in the context of multi-exponential extrapolation. 

One combination of error mitigation techniques that we have not explored here is pairing symmetry verification with error extrapolation without using quasi-probability. The naive version of such a combination is discussed in Ref.~\cite{caiResourceEstimationQuantum2020}. To make use of the results in this Article, one possible way is to approximate all the local error channels as the compositions of detectable and undetectable error channels, so that we can deal with them separately using hyperbolic extrapolation and exponential extrapolation. It would be very interesting to see the implementation details of such a method and how it compares to pure error extrapolation. 

We have only considered Pauli noise in this Article, thus it will also be interesting to see whether our arguments can be extended to other error channels like amplitude damping or coherent errors. In practice, we can transform any error channels into Pauli channels using Pauli twirling~\cite{wallmanNoiseTailoringScalable2016, caiConstructingSmallerPauli2019} and then apply our methods. Note that we can even perform further twirling like Clifford twirling to transform the error channels into group channels, which can be better mitigated as we have observed. Ways to transform a given error channel into a group channel can be an interesting area of investigation.

\section*{Data Availability}
The data that support the findings of this study are available from the corresponding author upon reasonable request.

\section*{Code Availability}
The code used in the current study are available from the corresponding author upon reasonable request.

\section*{Acknowledgements}
The author would like to thank Ying Li and Simon Benjamin for reading through the manuscript and providing valuable insights.

The author is supported by the Junior Research Fellowship from St John’s College, Oxford and acknowledges support from Quantum Motion Technologies Ltd and the QCS Hub (EP/T001062/1).

\section*{Author Contributions}
ZC is the sole author of the article.

\section*{Competing Interests}
The authors declare no conflict of interest.

\newpage
\appendix
\section{Post-processing Verification}\label{sec:post_proc_proj}
As discussed in \cref{sec:sym_proj}, when we cannot obtain the observable $O$ and the symmetry $S$ in the same run, we need to instead measure the Pauli components of $O\Pi_s$~\cite{bonet-monroigLowcostErrorMitigation2018}:
\begin{align*}
O\Pi_s = \sum_{k = 0}^{K-1} \alpha_k G_k = A \sum_{k = 0}^{K-1} \text{sgn}(\alpha_k) \frac{\abs{\alpha_k}}{A} G_k
\end{align*}
in which $A = \sum_{k = 0}^{K-1} \abs{\alpha_k}$. Note that $\alpha_k$ is real since $\Pi_s$ is Hermitian. When we run the circuit, we will perform $G_k$ measurement with $\frac{\abs{\alpha_k}}{A}$ probability and the measurement result will be multiplied with the sign factor $\text{sgn}(\alpha_k)$. The expectation value of such a sampling scheme scaled by $A$ will be $\expval{\Pi_s O}$. 

In the case that we cannot measure $S$ directly, then we also need to measured $\Pi_s$ using a similar scheme by breaking $\Pi_s$ into Pauli basis. 

For the case of a Pauli observable $O$, to perform symmetry verification in this way will require a sampling cost factor of~\cite{hugginsEfficientNoiseResilient2021}
\begin{align*}
C_{S 2} \sim \frac{1}{\Tr(\Pi_s \rho)^2}.
\end{align*}
which is $C_{S}^2$, i.e. we need to square the sampling cost to overcome our limitation of unable to measure our observables and symmetries in the same run. 

\section{Composition of Group Channels}\label{sec:prop_group_channel}
The group error $\mathcal{J}_{p, \mathbb{E}}$ of the group $\mathbb{E}$ with an error probability $p$ is defined in \cref{sec:group_channel}. Using $\widetilde{\mathbb{E}}$ to denote the set of generators of the group $\mathbb{E}$, we can rewrite the pure group error $\mathcal{J}_{1, \mathbb{E}}$ as:
\begin{align*}
\mathcal{J}_{1, \mathbb{E}} = \frac{1}{\abs{\mathbb{E}}} \sum_{E \in \mathbb{E}} \supop{E} =  \prod_{\widetilde{E} \in \widetilde{\mathbb{E}}} \frac{\supop{I} + \supop{\widetilde{E}}}{2}
\end{align*}
Using 
\begin{align*}
\left(\frac{\supop{I} + \supop{\widetilde{E}}}{2}\right) \left(\frac{\supop{I} + \supop{\widetilde{E}}}{2}\right) = \frac{\supop{I} + \supop{\widetilde{E}}}{2}
\end{align*}
we have:
\begin{align}
\mathcal{J}_{1,\mathbb{E}} \mathcal{J}_{1,\mathbb{B}} &= \left(\prod_{\widetilde{E} \in \widetilde{\mathbb{E}}} \frac{\supop{I} + \supop{\widetilde{E}}}{2}\right) \left(\prod_{\widetilde{B} \in \widetilde{\mathbb{B}}} \frac{\supop{I} + \supop{\widetilde{B}}}{2}\right)\nonumber\\
& = \prod_{\widetilde{D} \in \widetilde{\mathbb{E}} \cup \widetilde{\mathbb{B}}} \frac{\supop{I} + \supop{\widetilde{D}}}{2}\nonumber\\
& = \mathcal{J}_{1,\mathbb{D}} \label{eqn:group_channel_gen}
\end{align}
where $\mathbb{D}$ is the group generated by the union of of the generators of $\mathbb{E}$ and $\mathbb{B}$:
\begin{align*}
\widetilde{\mathbb{D}} = \widetilde{\mathbb{E}} \cup \widetilde{\mathbb{B}}.
\end{align*}
i.e. $\mathbb{D}$ is the group of elements that one can obtained by composing the elements in $\mathbb{E}$ and $\mathbb{B}$.

In the case of $\widetilde{\mathbb{E}}$ and $\mathbb{B}$ have no overlaps, i.e. $\widetilde{\mathbb{E}} \cap \mathbb{B} = 0$ (note not $\widetilde{\mathbb{B}}$ here since there is a degree of freedom in choosing $\widetilde{\mathbb{B}}$), we have $\widetilde{\mathbb{D}} = \widetilde{\mathbb{E}} + \widetilde{\mathbb{B}}$. We will have the same result if we have $\mathbb{E} \cap \widetilde{\mathbb{B}} = 0$ instead.  In such as case, for the subgroup $\mathbb{E}$  of $\mathbb{D}$, $\mathbb{B}$ is the corresponding quotient group, and vice versa. 

In the case of $\mathbb{E}$ is a subgroup of $\mathbb{B}$, we have
\begin{align}\label{eqn:idem_group_chan}
\mathcal{J}_{1,\mathbb{E}}\mathcal{J}_{1,\mathbb{B}} = \mathcal{J}_{1,\mathbb{B}}
\end{align} 
which leads to
\begin{align}\label{eqn:sink_group_chan}
\mathcal{J}_{p,\mathbb{E}}\mathcal{J}_{1,\mathbb{B}} = \mathcal{J}_{1,\mathbb{B}} \quad \forall p\text{ and }\mathbb{E} \subseteq \mathbb{B}.
\end{align}

Using \cref{eqn:sink_group_chan} we then have:
\begin{align*}
\mathcal{J}_{p, \mathbb{E}} \mathcal{J}_{q, \mathbb{E}} &= (1-q)\mathcal{J}_{p, \mathbb{E}} + q\mathcal{J}_{1, \mathbb{E}}\\
& = (1-q)(1-p)\mathcal{I} + \left(q + p - pq\right)\mathcal{J}_{1, \mathbb{E}}\\
& = \mathcal{J}_{q + p - pq, \mathbb{E}}
\end{align*}
which is still a group error of the same group $\mathbb{E}$ with a modified error probability.

For $\mathcal{J}_{q, \mathbb{E}}$ to be the inverse of $\mathcal{J}_{p, \mathbb{E}}$, we require:
\begin{align*}
q + p - pq &= 0\\
q &= -\frac{p}{1-p}.
\end{align*}
Hence,
\begin{align}\label{eqn:group_inverse_1}
\mathcal{J}_{p, \mathbb{E}}^{-1} = \mathcal{J}_{-\frac{p}{1-p}, \mathbb{E}}
\end{align}

\section{Removing Subgroup Components in a Group Channel}\label{sec:channel_after_qua}
\subsection{Form of the quasi-probability channel}
For a group error channel of the group $\mathbb{E}$, we would want to remove the error components $\mathbb{Q}$ in the channel where $\mathbb{Q}$ is a subgroup of $\mathbb{E}$. i.e. 
Thus we want to transform the channel:
\begin{align*}
\mathcal{J}_{p, \mathbb{E}} &= (1- p) \mathcal{I} + p\mathcal{J}_{1, \mathbb{E}}
\end{align*}
into the channel:
\begin{align*}
\mathcal{V}_q &= (1- q) \mathcal{I} + q \mathcal{V}_1
\end{align*}
where
\begin{equation}
\begin{aligned}\label{eqn:V_1}
\mathcal{V}_1 & = \frac{1}{\abs{\mathbb{E}} - \abs{\mathbb{Q}}} \sum_{V \in \mathbb{E}, V \not \in \mathbb{Q}}  \supop{V}\\
&= \frac{1}{\abs{\mathbb{E}} - \abs{\mathbb{Q}}}\left(\abs{\mathbb{E}}\mathcal{J}_{1, \mathbb{E}} - \abs{\mathbb{Q}}\mathcal{J}_{1, \mathbb{Q}}\right).
\end{aligned}
\end{equation}

Using \cref{eqn:idem_group_chan}, we have 
\begin{align*}
\mathcal{V}_q \mathcal{J}_{1, \mathbb{E}} = \mathcal{J}_{1, \mathbb{E}},
\end{align*}
along with \cref{eqn:group_inverse_1}, the channel we need to perform to transform $\mathcal{J}_{p, \mathbb{E}}$ to $\mathcal{V}$ will be:
\begin{align}
\mathcal{V}_{q} \mathcal{J}_{p, \mathbb{E}}^{-1}  &= \mathcal{V}_{q} \mathcal{J}_{\frac{p}{p-1}, \mathbb{E}}= \frac{1}{1-p}\left( \mathcal{V}_q - p\mathcal{J}_{1, \mathbb{E}}\right)\nonumber\\
& = \frac{1}{1-p} \bigg[\left(1- q\right)\mathcal{I} + q \mathcal{V}_1 - p\mathcal{J}_{1, \mathbb{E}}\bigg]. \label{eqn:chan_to_apply}
\end{align} 
Using \cref{eqn:V_1}, we have:
\begin{align*}
\mathcal{J}_{1, \mathbb{E}} = \frac{\abs{\mathbb{E}} - \abs{\mathbb{Q}}}{\abs{\mathbb{E}}}\mathcal{V}_1 + \frac{\abs{\mathbb{Q}}}{\abs{\mathbb{E}}}\mathcal{J}_{1, \mathbb{Q}}.
\end{align*}
It turns \cref{eqn:chan_to_apply} into:
\begin{equation}\label{eqn:chan_to_apply_2}
\begin{aligned}
\mathcal{V}_{q} \mathcal{J}_{p, \mathbb{E}}^{-1} &= \frac{1}{1-p} \bigg[\left(1- q\right)\mathcal{I} + \left(q - p_d\right) \mathcal{V}_1 - \frac{\abs{\mathbb{Q}}}{\abs{\mathbb{E}}}p \mathcal{J}_{1, \mathbb{Q}}\bigg]\\
& = \frac{1}{1-p} \bigg[\left(1- q\right)\supop{I}  + \frac{q - p_d}{\abs{\mathbb{E}} - \abs{\mathbb{Q}}} \sum_{\substack{V \in \mathbb{E}\\ V \not \in \mathbb{Q}}}  \supop{V} - \frac{p}{\abs{\mathbb{E}}}\sum_{Q \in \mathbb{Q}}  \supop{Q}\bigg].
\end{aligned}
\end{equation}
in which $p_d = \frac{\abs{\mathbb{E}} - \abs{\mathbb{Q}}}{\abs{\mathbb{E}}}p$ as defined in \cref{eqn:prob_detected}, which is the probability of the errors in $V \in \mathbb{E} - \mathbb{Q}$ occurring.

At $q = p_d$, the channel we need to apply is just:
\begin{equation}
\begin{aligned}
\mathcal{V}_{p_d} \mathcal{J}_{p, \mathbb{E}}^{-1} = \frac{1}{1-p} \bigg[\left(1- p_d\right)\mathcal{I} - \frac{\abs{\mathbb{Q}}}{\abs{\mathbb{E}}}p \mathcal{J}_{1, \mathbb{Q}}\bigg].
\end{aligned}
\end{equation}
i.e. we only need to apply components in $\mathbb{Q}$ in our quasi-probability channel.

And using \cref{eqn:V_1}, our resultant channel is simply:
\begin{align*}
\mathcal{V}_{p_d} &= (1- p_d) \mathcal{I} + p_d \mathcal{V}_1\\
& = (1- p_d) \supop{I} +  \frac{p_d}{\abs{\mathbb{E}} - \abs{\mathbb{Q}}} \sum_{V \in \mathbb{E}, V \not \in \mathbb{Q}}  \supop{V}.
\end{align*}
same as what we derived in \cref{sec:comb_ver_qua}.

\subsection{Cost of the quasi-probability channel}
Decomposing the quasi-probability channel that we need to implement into its components:
\begin{align*}
\mathcal{V}_q \mathcal{J}_{p, \mathbb{E}}^{-1} = \sum_{E \in \mathbb{E}} \beta_E \supop{E}
\end{align*}
and compared to \cref{eqn:chan_to_apply_2}, we have:
\begin{align*}
\beta_Q &= - \frac{p}{\left(1-p\right)\abs{\mathbb{E}}} \leq 0 \quad \forall Q \in \mathbb{Q}, Q \neq I\\
\beta_V &= \frac{1}{1-p}\left(\frac{q - p_d}{\abs{\mathbb{E}} - \abs{\mathbb{Q}}}\right) \quad \forall V \in \mathbb{E}, V \not \in \mathbb{Q}.
\end{align*}
The normalisation factor of implementing the quasi-probability channel $\mathcal{V} \mathcal{J}_{p, \mathbb{E}}^{-1}$ is:
\begin{align}\label{eqn:quasi_norm}
B &= \sum_{E \in  \mathbb{E}} \abs{\beta_E}=  1 + 2\left(\sum_{\beta_E < 0} \abs{\beta_E}\right)
\end{align}
where we have used $\sum_{E \in  \mathbb{E}} \beta_E = 1$. 

It means that the cost of implementing the quasi-probability channel is split into two case by the threshold error rate $q = p_d$ of the final channel $\mathcal{V}_{q}$.
\subsubsection*{Targeted error rate above or at the threshold}
This means:
\begin{align*}
q \geq p_d \quad \Rightarrow \quad \beta_V \geq 0.
\end{align*}
Now using \cref{eqn:quasi_norm}, we have
\begin{align*}
B_1 & = 1 + 2 \left(\abs{\mathbb{Q}} - 1\right) \abs{\beta_Q}\\
& = 1 + 2\frac{\left(\abs{\mathbb{Q}} - 1\right)p}{\abs{\mathbb{E}}\left(1- p\right)}.
\end{align*}
The cost to implementing the transformation channel $\mathcal{V}_q \mathcal{J}_{p, \mathbb{E}}^{-1}$ is then
\begin{align*}
C_{Q1, q} & = B_1^2 \approx 1 + 4\frac{\left(\abs{\mathbb{Q}} - 1\right)p}{\abs{\mathbb{E}}} + \mathcal{O}(p^2).
\end{align*}
i.e. the cost to implementing the quasi probability channel in this regime is independent of $q$. However, the exact quasi-probability channel we need to implement will still change with $q$. 

Uses \cref{eqn:p_e} and \cref{eqn:prob_detected}, we have:
\begin{align*}
C_{Q1, q} & = 1 + 4\left(p_{\epsilon} - p_d\right).
\end{align*}

\subsubsection*{Target error rate below the threshold} 
This means:
\begin{align*}
q < p_d \quad \Rightarrow \quad \beta_V < 0
\end{align*}
Now using \cref{eqn:quasi_norm}, we have
\begin{align*}
B_2 & = 1 + 2 \left(\abs{\mathbb{Q}} - 1\right) \abs{\beta_Q} + 2 \left(\abs{\mathbb{E}} - \abs{\mathbb{Q}}\right) \abs{\beta_V}\\
& =  B_1 + \frac{2}{1-p}\left(\frac{\abs{\mathbb{E}} - \abs{\mathbb{Q}}}{\abs{\mathbb{E}}}p - q\right)\\
& =  1 + 2\frac{\left(\abs{\mathbb{E}} - 1\right)p}{\abs{\mathbb{E}}\left(1- p\right)} - \frac{2q}{1-p}
\end{align*}
The cost to implementing the transformation channel $\mathcal{V}_q \mathcal{J}_{p, \mathbb{E}}^{-1}$ is just
\begin{align*}
C_{Q1, q}  = B_2^2 &= 1 + 4\left(\frac{\abs{\mathbb{E}} - 1}{\abs{\mathbb{E}}}p - q \right) + \mathcal{O}(p^2) + \mathcal{O}(pq)\\
& \approx 1+ 4\left(p_\epsilon - q\right)
\end{align*}
where we have use the expression of $p_\epsilon$ in \cref{eqn:p_e}. Here $1+ 4p_\epsilon$ is the cost to invert the entire channel $\mathcal{J}_{p, \mathbb{E}}$. For each bit of remaining error probability $q$ in our final channel $\mathcal{V}_q$, we can reduce the cost factor by $4q$ until we hit the threshold $q = p_d$. This is the same as what we obtained in \cref{eqn:pauli_trans_cost}.

\section{Decay of Pauli Expectation Value under Group Channels}\label{sec:der_exp_decay}
\subsection{Overall derivation}
In \cref{sec:err_extrapolation}, we are looking at a circuit of the form $U = \prod_{m = M}^{1} V_m$ with each gate decomposed into their Pauli components: $V_m = \sum_{j_{m}} \alpha_{mj_{m}} G_{mj_m}$.

Now if a pure group error $\mathcal{J}_{1, \mathbb{E}_f}$ happens between $V_{f}$ and $V_{f+1}$, and we denote 
\begin{align*}
\{a:b\} &= \{a, a+1, \cdots b\}\\
U_{\mathbb{M}} = \prod_{m \in \mathbb{M}} V_m,\quad
G_{\vec{j}_{\mathbb{M}}} &= \prod_{m \in \mathbb{M}}  G_{mj_m},\quad
\alpha_{\vec{j}_{\mathbb{M}}} = \prod_{m \in \mathbb{M}} \alpha_{mj_{m}},
\end{align*}
then the expectation value of a Pauli observable $O$ is:
\begin{align*}
&\quad \Tr(U_{f+1:M}\mathcal{J}_{1, \mathbb{E}_f}(U_{1:f}\rho U_{1:f}^\dagger) U_{f+1:M}^\dagger O) \\
&= \sum_{\vec{i}, \vec{j}} \alpha_{\vec{i}}^* \alpha_{\vec{j}} \Tr(G_{\vec{j}_{f+1:M}}\mathcal{J}_{1, \mathbb{E}_f}(G_{\vec{j}_{1:f}}\rho G_{\vec{i}_{1:f}}^\dagger) G_{\vec{i}_{f+1:M}}^\dagger O)\\
&= \sum_{\vec{i}, \vec{j}} X_f(\vec{i}, \vec{j})  \alpha_{\vec{i}}^* \alpha_{\vec{j}} \Tr(\rho G_{\vec{i}}^\dagger OG_{\vec{j}})
\end{align*}
where in the last step we have use \cref{eqn:effect_group_ch} and we have defined:
\begin{align*}
X_f(\vec{i}, \vec{j}) = \frac{1}{2^{\abs{\widetilde{\mathbb{E}}_f}}}\prod_{\widetilde{E} \in \widetilde{\mathbb{E}}_f} \left(1 + \eta(\widetilde{E}, G_{\vec{i}_{f+1:M}}^\dagger OG_{\vec{j}_{f+1:M}})\right).
\end{align*}
Here $X_f(\vec{i}, \vec{j})$ takes the value $0$ if the effective observable $G_{\vec{i}_{f+1:M}}^\dagger OG_{\vec{j}_{f+1:M}}$ right after the noise does not commute with all elements in $\mathbb{E}_f$, and takes the value $1$ otherwise. For the former case, we will say the information about $G_{\vec{i}}^\dagger OG_{\vec{j}}$ is ``erased'' by the error at location $f$. Note that the effect of the noise does not relate to the gates implemented before it at all. For a given effective observable $G_{\vec{i}}^\dagger OG_{\vec{j}}$, the fraction of error locations that can ``erase'' its information is simply:
\begin{align*}
    \gamma_{\vec{i}, \vec{j}} = 1 - \frac{1}{M}\sum_{f = 1}^{M}  X_f(\vec{i}, \vec{j}).
\end{align*}

Denoting $\expval{O_{\mathbb{L}}}$ as the expectation value we obtain when the set of error location is $\mathbb{L}$, we have:
\begin{align*}
\expval{O_{\mathbb{L}}} & = \sum_{\vec{i}, \vec{j}} \left(\prod_{f \in \mathbb{L}}X_f(\vec{i}, \vec{j}) \right) \alpha_{\vec{i}}^* \alpha_{\vec{j}} \Tr(\rho G_{\vec{i}}^\dagger OG_{\vec{j}}).
\end{align*}
Recall that $\expval{O_{\abs{\mathbb{L}} = l}}$ is the expectation value we obtain when there are $l$ errors in the circuit, regardless of the error location. By definition we have:
\begin{align}\label{eqn:oEl}
&\expval{O_{\abs{\mathbb{L}} = l}} = \frac{1}{{}^MC_l}\sum_{\abs{\mathbb{L}} = l} \expval{O_{\mathbb{L}}}\nonumber\\
&=  \sum_{\vec{i}, \vec{j}} \left(\frac{1}{{}^MC_l}\sum_{\abs{\mathbb{L}} = l}\prod_{f \in \mathbb{L}}  X_f(\vec{i}, \vec{j})\right) \alpha_{\vec{i}}^* \alpha_{\vec{j}} \Tr(\rho G_{\vec{i}}^\dagger OG_{\vec{j}}).
\end{align}

As proven in \cref{sec:bern_sample}, in the limit of large $M$ and non-vanishing $1- \gamma_{\vec{i}, \vec{j}}$, we have:
\begin{align*}
\frac{1}{{}^MC_l}\sum_{\abs{\mathbb{L}} = l}\prod_{f \in \mathbb{L}}  X_f(\vec{i}, \vec{j}) \approx \left(\frac{1}{M}\sum_{f = 1}^{M}  X_f(\vec{i}, \vec{j})\right)^l =  \left(1-\gamma_{\vec{i}, \vec{j}}\right)^l.
\end{align*}
Thus \cref{eqn:oEl} can be approximated as:
\begin{align}
\expval{O_{\abs{\mathbb{L}} = l}} 
&\approx  \sum_{\vec{i}, \vec{j}} \left(1-\gamma_{\vec{i}, \vec{j}}\right)^l \alpha_{\vec{i}}^* \alpha_{\vec{j}} \Tr(\rho G_{\vec{i}}^\dagger OG_{\vec{j}}) \nonumber\\
&=  2\sum_{\vec{i} > \vec{j}} \left(1-\gamma_{\vec{i}, \vec{j}}\right)^l \Re{\alpha_{\vec{i}}^* \alpha_{\vec{j}} \Tr(\rho G_{\vec{i}}^\dagger OG_{\vec{j}})} \label{eqn:exp_group_decay}
\end{align}
where we have use $\gamma_{\vec{i}, \vec{j}} = \gamma_{\vec{j}, \vec{i}}$ from the definition of $\gamma_{\vec{i}, \vec{j}}$.

\subsection{Expansion of the sum of Bernoulli samples}\label{sec:bern_sample}

We denote $X_f$ as the $f$th sample from a Bernoulli distribution. From the total $M$ samples taken, we can estimate the success probability denoted $r$:
\begin{align*}
\frac{\sum_{f} X_f}{M} = r.
\end{align*}

Now we define
\begin{align*}
Y_l = \sum_{\abs{\mathbb{L}} = l } \prod_{f \in \mathbb{L}} X_f.
\end{align*}
Then $Y_1$ is just the sum of these samples:
\begin{align*}
Y_1 = \sum_{f} X_f = Mr.
\end{align*}
Using multinomial expansion, we have
\begin{align*}
Y_1^l = \left(\sum_{f=1}^{M} X_f\right)^l = \sum_{\sum_{f}n_f = l} \begin{pmatrix}
l\\n_1,n_2,\cdots,n_M
\end{pmatrix}\prod_{f=1}^{M}X_{f}^{n_f}
\end{align*}
where the multinomial coefficient is
\begin{align*}
\begin{pmatrix}
l\\n_1,n_2,\cdots,n_M
\end{pmatrix} = \frac{l!}{\prod_{f=1}^{M}n_f!}.
\end{align*}
It is the coefficient of the term $\prod_{f=1}^{M}X_{f}^{n_f}$, which is the number of ways the distribute $l$ distinct balls (the total power of $l$) into $M$ distinct bins ($M$ terms of $X_{f}$ for $f \in \{1, 2, 3,\cdots, M\}$) such that the number of balls in bin $f$ is $n_f$ (the power of $X_{f}$ is $n_f$).

Now using $X_f^n = X_f$ for any $n$, we have 
\begin{align*}
\left(\sum_{f=1}^{M} X_f\right)^l &= \sum_{b = 1}^{l}\begin{Bmatrix}l\\b\end{Bmatrix}b! \sum_{\abs{\mathbb{L}} = b } \prod_{f \in \mathbb{L}} X_f\\
Y_1^l &= \sum_{b = 1}^{l}\begin{Bmatrix}l\\b\end{Bmatrix}b! Y_b
\end{align*}
where $\mathbb{L}$ is the subset of non-empty bins and $\begin{Bmatrix}l\\b\end{Bmatrix}$ is the Stirling number of the second kind, which is the number of ways to distribute $l$ \emph{distinct} balls into these $b$ \emph{identical} bins such that none are empty. And $\begin{Bmatrix}l\\b\end{Bmatrix}b!$ is just the number of ways of distribute $l$ \emph{distinct} balls into these $b$ \emph{distinct} bins such that none are empty.

Hence, 
\begin{align}
\begin{Bmatrix}l\\l\end{Bmatrix}l!Y_l &= Y_1^l - \sum_{b = 1}^{l-1}\begin{Bmatrix}l\\b\end{Bmatrix}b! Y_b \nonumber\\
l!Y_l &= Y_1^l - \sum_{b = 1}^{l-1}\begin{Bmatrix}l\\b\end{Bmatrix}b! Y_b\label{eqn:Y_expansion}
\end{align}
where we have used $\begin{Bmatrix}l\\l\end{Bmatrix} = 1$. 

Hence, in the limit of large $M$ and assuming the success sample fraction $r$ is non-vanishing, we have:
\begin{align*}
\frac{Y_l}{{}^MC_l} \approx \frac{l!Y_l}{M^l} \approx \frac{Y_1^l +  \mathcal{O}(Y_1^{l-1})}{M^l} = r^l + \mathcal{O}(M^{-1}).
\end{align*}
From the definition of $Y_l$ we have:
\begin{align*}
\frac{1}{{}^MC_l}\sum_{\abs{\mathbb{L}} = l } \prod_{f \in \mathbb{L}} X_f \approx r^l + \mathcal{O}(M^{-1}).
\end{align*}
The estimated failure probability of the Bernoulli distribution is:
\begin{align*}
    \gamma = 1-r.
\end{align*}
which gives:
\begin{align*}
    \frac{1}{{}^MC_l}\sum_{\abs{\mathbb{L}} = l } \prod_{f \in \mathbb{L}} X_f \approx \left(1-\gamma\right)^l + \mathcal{O}(M^{-1}).
\end{align*}

\section{Decay of Pauli Expectation Value under Pauli Channels}\label{sec:pauli_expec_decay}
Any Pauli channel can be decomposed into a set of group channels basis:
\begin{align}\label{eqn:Pauli_group_decomp}
\mathcal{P}_p = \left(1 - p\right) \mathcal{I} + p \sum_{j} \beta_j \mathcal{J}_{1, \mathbb{E}_j}
\end{align}
where $\sum_j \beta_j = 1$. In the most naive way, we can have $\mathbb{E}_j = \{I, G_j\}$ for all $G_j$ in the set of Pauli operators and $\frac{p\beta_j}{2}$ being the error rate of the Pauli error $G_j$. 

For a circuit with $M$ Pauli error locations, different error locations might experience different Pauli noise of different strengths. We will denote the union of the group channel basis needed to describe all of these Pauli channels as $\{\mathcal{J}_{1, \mathbb{E}_j}\ |\ 1 \leq j \leq J\}$. Now we can split each Pauli error location into $J$ group error locations, with the $j$th location can only have the group error $\mathcal{J}_{1, \mathbb{E}_j}$ occurring. Hence, we have in total $MJ$ group error locations. Using the same arguments in \cref{sec:der_exp_decay}, we can obtain the same equation as \cref{eqn:exp_group_decay}, which will lead to \cref{eqn:pauli_expect_decay}:
\begin{align}\label{eqn:Pauli_Ol}
 \expval{O_{\abs{\mathbb{L}} = l}} & = \sum_{k = 1}^{K} A_k \left(1-\gamma_k\right)^l,
\end{align}
but now $\abs{\mathbb{L}}$, $l$ and $\gamma_{k}$ are defined in terms of \emph{group} error locations instead of simply error locations.

Following the same arguments in \cref{sec:nisq_limit}, but focusing on group errors instead of simply errors. When a Pauli channel is written in the form of \cref{eqn:Pauli_group_decomp}, the error rate $p$ defined in this way is the probability that one group error occurs (which could be any one of the basis group errors.). i.e. the number of group error at each Pauli error location is a Bernoulli variable with the success probability $p$. The mean circuit \emph{group} error count is just the sum of the group error probability of all the error locations, which we will denote as $\mu$. Again taking the NISQ limit and using the Le Cam's theorem, the number of \emph{group} errors occurring in the circuit will follow a Poisson distribution with the mean $\mu$, which is just \cref{eqn:poisson_prob}:
\begin{align}
P_l = e^{-\mu}\frac{\mu^l}{l!}.
\end{align}
Combining with \cref{eqn:Pauli_Ol}, we can again obtain the exponential decay of the expectation value with increase of mean circuit (group) error count $\mu$ just like in \cref{sec:err_extrapolation}. 

\section{Quasi-probability with Symmetry Verification}\label{sec:comb_ver_qua}
After using quasi-probability to remove all the local undetectable errors in the circuit, which reduces the mean circuit error count from $\mu$ to $\mu_d$ as discussed in \cref{sec:nisq_limit}, we can further suppress the remaining errors to achieve a mean circuit error count of $\nu \leq \mu_d$ and follow the same arguments in \cref{sec:nisq_limit} with $\nu$ in place of $\mu_d$. Using \cref{eqn:partial_qua_cost}, the sampling cost factor of the required quasi-probability transformation is: 
\begin{align}\label{eqn:cost_qua_for_comb}
    C_{Q, \nu} \approx e^{4\left(\mu_{\epsilon} - \nu\right)}\quad \nu \leq \mu_d
\end{align}
where we have made use of $\nu_\epsilon = \nu$ for the detectable error channels like $\mathcal{V}_q$ in \cref{eqn:quasi_final_arbit_channel}.

In the resultant circuit with only local detectable errors present, the sampling cost of applying symmetry verification is given by \cref{eqn:nisq_sym_cost} with $\nu$ in place of $\mu_d$:
\begin{align*}
    C_{S, \nu} = \frac{1}{e^{-\nu} \cosh(\nu)}.
\end{align*}
Hence, the total sampling cost including quasi-probability is:
\begin{align*}
    C_{QS, \nu} = C_{Q, \nu}C_{S, \nu}&= \frac{e^{4\mu_{\epsilon}}}{e^{3\nu}\cosh(\nu)}
\end{align*}
which is smaller than the cost of using quasi-probability to remove all of our errors without the help of symmetry verification in \cref{eqn:pure_qua_cost}: $C_{Q, 0} \approx e^{4\mu_\epsilon}$. 

However, we need to note that with the pure quasi-probability method, we can remove all of the noise, while when we combine with symmetry verification, the fraction of erroneous circuit runs after applying our combined error mitigation is given by \cref{eqn:circ_err_after_sym} with $\nu$ in place of $\mu_d$:
\begin{align*}
    P_{circ} &= \frac{1}{2} \left(1 - e^{-\nu}\right)^2.
\end{align*}
Hence, we must choose a $\nu$ such that $P_{circ}$ is the circuit error rate that we can tolerate. 

The saving of sampling cost when combining symmetry verification with quasi-probability over pure quasi-probability can be written as:
\begin{align*}
    \frac{C_{Q, 0}}{C_{QS, \nu}} &= e^{3\nu}\cosh(\nu)\\
    & = \left(1 - \sqrt{2P_{circ}}\right)^{-4} \left(1 - \sqrt{2P_{circ}} + P_{circ}\right).
\end{align*}
To achieve $P_{circ} \ll 1$, we have:
\begin{align*}
    \frac{C_{Q, 0}}{C_{QS, \nu}} &\approx 1 + 3 \sqrt{2P_{circ}} + \mathcal{O}(P_{circ}).
\end{align*}

Therefore by combining with symmetry verification, the factor of saving in the sampling cost over pure quasi-probability is $\nicefrac{C_{Q, 0}}{C_{QS, \nu}} = 1.5$ for the circuit error rate $P_{circ} = 10^{-2}$ and $\nicefrac{C_{Q, 0}}{C_{QS, \nu}} = 1.15$ for the circuit error rate $P_{circ} = 10^{-3}$. To push the circuit error rate any lower, the saving in the sampling cost will be negligible. And in the limit of $P_{circ} = 0 \Rightarrow \nu = 0$, we just have pure quasi-probability which removes all the errors.

We can of course apply hyperbolic extrapolation on top of this, which is the same as \cref{sec:comb_err_miti}, but with $\nu$ in place of $\mu_d$. The cost function is thus similar to \cref{eqn:comb_opt_cost}
\begin{align*}
    C_{QH, \nu}(\gamma) = e^{4\mu_\epsilon} \frac{\cosh(\nu)\cosh(2\left(1- \gamma\right) \nu)}{e^{3\nu}}.
\end{align*}
We need to note that the cost expression here is only valid for $0 \leq \nu \leq \mu_d$ following from \cref{eqn:cost_qua_for_comb}, thus the minimum cost that we can achieve will be at $\nu = \mu_d$. The only reason to try to push $\nu$ below $\mu_d$ is when such an action will result in easier (smaller $K$) and/or better fitting to our extrapolation curves.

\section{Cost of Error Extrapolation}
\subsection{Cost for two-point extrapolation} \label{sec:cost_two_pt_extr}
Suppose for an observable $O$, we can estimate its expectation value by combining the expectation value of two other observables $A$ and $B$. Then by denoting the estimate as $\expval{O_0}$, we have:
\begin{align}
\expval{O} \approx \expval{O_0} := f(\expval{A}, \expval{B})
\end{align}
for some estimation function $f$.

Now suppose we take $N$ samples in total and $\alpha$ is the fraction of $A$ samples within, then using $\overline{A}$, $\overline{B}$ to denote the sample averages, we can obtain the sample estimate of $\expval{O_0}$ (and thus $\expval{O}$):
\begin{align}
\overline{O}_0 := f(\overline{A}, \overline{B}).
\end{align}
Note that we have abuse the notation here since $\expval{O_{0}}$ and $\overline{O}_{0}$ are \emph{not} the expectation value and the sample average of some observable $O_{0}$. There is no such an observable. Rather, $\overline{O}_{0}$ is \emph{exactly defined} as the estimate of $\expval{O}$ after $N$ total samples of $A$ and $B$ using the equation above, and similarly $\expval{O_0}$ is defined as the case $N \rightarrow \infty$. Note that $O_{0} \not \equiv O$ as well since $O$ is an actual observable of the noiseless computation that lead us to $\expval{O}$.

We will try to compare the variance of the sample estimate $\overline{O}_0$ against the variance of the noiseless sample average $\overline{O}$. The variances of various sampling averages follow these equations:
\begin{align*}
\var{\overline{A}}  &= \frac{\var{A}}{\alpha N}\\
\var{\overline{B}}  &= \frac{\var{B}}{(1-\alpha)N}\\
\var{\overline{O}}  &= \frac{\var{O}}{N}.
\end{align*}
Hence, assuming $\var{A} \approx \var{B} \approx \var{O}$, the variance of the sample estimate is:
\begin{align*}
\var{\overline{O}_0} & = \left(\pdv{\overline{O}_0}{\overline{A}}\right)^2\var{\overline{A}} + \left(\pdv{\overline{O}_0}{\overline{B}}\right)^2\var{\overline{B}}\\
& = \left(\left(\pdv{\overline{O}_0}{\overline{A}}\right)^2\frac{1}{\alpha} + \left(\pdv{\overline{O}_0}{\overline{B}}\right)^2\frac{1}{\left(1 - \alpha\right)} \right) \var{\overline{O}}.
\end{align*}

Hence, when sampling for $\overline{O}_0$ instead of $\overline{O}$, the additional cost factor is:
\begin{align}\label{eqn:cost_frac}
C(\alpha) = \frac{a^2}{\alpha} + \frac{b^2}{1-\alpha}
\end{align}
with 
\begin{align*}
a = \abs{\pdv{\overline{O}_0}{\overline{A}}},\quad b = \abs{\pdv{\overline{O}_0}{\overline{B}}}
\end{align*}

We then have the extremal point being
\begin{align*}
\alpha_{\pm} = \frac{a}{a \pm b}.
\end{align*}
Since $\alpha_- = \frac{a}{a - b} \geq 1$, we will only keep $\alpha_+$. We can also obtain $C''(\alpha_+) \geq 0$, which means that it is a local minimum, whose value is
\begin{align}\label{eqn:min_cost_extr}
C(\alpha_+) = \left(a + b\right)^2.
\end{align}
i.e. the minimal cost factor is achieved when the fraction of $A$ samples is $\alpha_+$.

For the naive version of evenly distributing all the samples: $\alpha = 0.5$, we have
\begin{align} \label{eqn:even_cost_extr}
C(0.5)  = 2 (a^2 + b^2).
\end{align}
When compared to the optimal distribution, we have:
\begin{align*}
C(\alpha_+) \leq C(0.5)  = 2 (a^2 + b^2) \leq 2 \left(a + b\right)^2 = 2 C(\alpha_+).
\end{align*}
i.e. the saving in the number of samples by using optimal sample distribution will be less than half. The saving will be exactly half in the case of $a \gg b$ or $b \gg a$. In practice, $a$ and $b$ is often unknown and thus it is hard to achieve the optimal sample distribution. Hence, in most of the cases in this Article, we just use the naive sample distribution, which should be of the same order of magnitude as the optimal case.

\subsection{Cost of exponential extrapolation} \label{sec:extr_cost_detail}
For an observable $O$, we can estimate its expectation value by probing at the error rate $\mu$ and $\lambda \mu$ and fitting with an exponential curve to get the zero-noise value:
\begin{align}\label{eqn:exp_extrapolate_1}
\expval{O} \approx \expval{O_0} := \left(\frac{\expval{O_{\mu}}^\lambda}{\expval{O_{\lambda \mu}} }\right)^{\frac{1}{\lambda - 1}}
\end{align}
Note that $\expval{O}$ will only be exactly the same as $\expval{O_0}$ if the observable follows strictly a single exponential decay with increased noise.

Now following \cref{sec:cost_two_pt_extr} with $A = O_\mu$, $B = O_{\lambda\mu}$ and using $\overline{O}_\mu \approx \overline{O}_0 e^{-\gamma \mu}$, we have:
\begin{equation}\label{eqn:pdv_exp_extr}
    \begin{aligned}
    a &= \abs{\pdv{\overline{O}_0}{\overline{O}_\mu} } =  \frac{\lambda}{\lambda - 1}\left(\frac{\overline{O}_{\mu}}{\overline{O}_{\lambda \mu} }\right)^{\frac{1}{\lambda - 1}} = \frac{\lambda}{\lambda - 1} e^{\gamma \mu}\\
    b &= \abs{\pdv{\overline{O}_0}{\overline{O}_{\lambda\mu}}}  =   \frac{1}{\lambda - 1}\left(\frac{\overline{O}_{\mu}}{\overline{O}_{\lambda \mu} }\right)^{\frac{\lambda}{\lambda - 1}} =  \frac{1}{\lambda - 1} e^{ \lambda \gamma \mu}.
    \end{aligned}
\end{equation}
Hence, the sampling cost factor of exponential extrapolation can be obtained using \cref{eqn:even_cost_extr}:
\begin{align}\label{eqn:exp_cost_sim} 
C_E = 2\left(a^2 + b^2\right) = 2\frac{\lambda^2 e^{2\gamma \mu} + e^{ 2\lambda \gamma \mu}}{\left(\lambda - 1\right)^2}.
\end{align}

\subsection{Cost of quasi-probability with exponential extrapolation}\label{sec:QE_cost}
As outlined in \cref{sec:comb_err_miti}, we use quasi-probability to suppress the error rate from $\mu$ to $\nu = \frac{\mu}{\lambda}$ and then use the point at $\mu = \lambda \nu$ and $\nu$ to perform exponential extrapolation. Thus we have the same equation as \cref{eqn:exp_extrapolate_1} with $\mu \rightarrow \nu$:
\begin{align}
\expval{O_0} = \left(\frac{\expval{O_{\nu}}^\lambda}{\expval{O_{\lambda \nu}} }\right)^{\frac{1}{\lambda - 1}}.
\end{align}
Here $\lambda \nu = \mu$ is the original error rate and thus $\expval{O_{\lambda \nu}}$ will be obtained via direct sampling, while $\nu$ is the quasi-probability suppressed error rate, thus $\expval{O_{\nu}}$ will be obtain via quasi-probability. Thus using \cref{eqn:qus_trans_var}, we have
\begin{align*}
\expval{O_{\nu}} = Q \expval{O_{Q}}.
\end{align*}

Now following \cref{sec:cost_two_pt_extr} with $A = O_Q$, $B = O_{\lambda\nu}$ and using \cref{eqn:pdv_exp_extr} with $\nu$ in place of $\mu$, we have:
\begin{align*}
a &= \abs{\pdv{\overline{O}_0}{\overline{O}_Q} } = \abs{\pdv{\overline{O}_0}{\overline{O}_\nu} \dv{\overline{O}_\nu}{\overline{O}_Q}} = \frac{\lambda}{\lambda - 1} e^{\gamma \nu} Q\\
b &= \abs{\pdv{\overline{O}_0}{\overline{O}_{\lambda\nu}}} =  \frac{1}{\lambda - 1} e^{ \lambda \gamma \nu}.
\end{align*}
Using \cref{eqn:quasi_cost}, we have $Q = \sqrt{C_{Q, \nu}}$. From \cref{eqn:partial_qua_cost}, we have $C_{Q, \nu} = e^{4\left(\mu_\epsilon - \nu_{\epsilon}\right)}$. Since in extrapolation, we need to suppress all error components evenly, we have $\mu_\epsilon  = \lambda \nu_{\epsilon}$ just like the relation between $\mu$ and $\nu$. Hence, we have:
\begin{align*}
C_{Q, \nu} &=  e^{4\left(\lambda - 1\right)\nu_\epsilon}\\
Q &= \sqrt{C_{Q, \nu}} =e^{2\left(\lambda - 1\right)\nu_\epsilon},
\end{align*}
which gives
\begin{align*}
a &= \abs{\pdv{\overline{O}_0}{\overline{O}_Q} } = \abs{\pdv{\overline{O}_0}{\overline{O}_\nu} \dv{\overline{O}_\nu}{\overline{O}_Q}} = \frac{\lambda}{\lambda - 1} e^{\gamma \nu + 2\left(\lambda - 1\right) \nu_\epsilon}\\
b &= \abs{\pdv{\overline{O}_0}{\overline{O}_{\lambda\nu}}} =  \frac{1}{\lambda - 1} e^{ \lambda \gamma \nu}.
\end{align*}
Hence, the sampling cost factor of quasi-probability with exponential extrapolation can be obtained using \cref{eqn:even_cost_extr}:
\begin{equation}
    \begin{aligned}
    C_{QE} = 2\left(a^2 + b^2\right) &= 2\frac{\lambda^2 e^{2\gamma \nu + 4\left(\lambda - 1\right) \nu_\epsilon} + e^{ 2\lambda \gamma \nu}}{\left(\lambda - 1\right)^2}\\
    &= 2\frac{\lambda^2 e^{\frac{2}{\lambda}\left[\gamma \mu + 2\left(\lambda - 1\right) \mu_\epsilon\right]} + e^{2\gamma \mu}}{\left(\lambda - 1\right)^2}.
    \end{aligned}
\end{equation}

\subsection{Cost of hyperbolic extrapolation}\label{sec:hyper_cost}
Performing error extrapolation using \cref{eqn:get_O0}, we have:
\begin{align*}
\expval{O_0} &= \text{sgn}\left(\expval{O_{c, \nu}}\right) \sqrt{\expval{O_{c, \nu}}^2\cosh[2](\nu) - \expval{O_{s, \nu}}^2\sinh[2](\nu)}.
\end{align*}
Following the arguments in \cref{sec:cost_two_pt_extr} with $A = O_{c, \nu}$, $B = O_{s, \nu}$, and using \cref{eqn:cosh_sinh}, we have
\begin{align*}
a = \abs{\pdv{\overline{O}_{0}}{\overline{O}_{c, \nu}}} &= \cosh[2](\nu)\frac{\overline{O}_{c, \nu}}{\overline{O}_{0}} = \cosh(\nu)\cosh(\left(1- \gamma\right) \nu) \\
b = \abs{\pdv{\overline{O}_{0}}{\overline{O}_{s, \nu}}} &= \sinh[2](\nu)\frac{\overline{O}_{s, \nu}}{\overline{O}_{0}} = \sinh(\nu)\sinh(\left(1- \gamma\right) \nu).
\end{align*}
The fraction of $O_{c, \nu}$ samples among all the samples can be obtained from \cref{eqn:exp_fraction}:
\begin{align*}
\alpha = e^{-\nu} \cosh(\nu).
\end{align*}
Hence, the cost factor can be obtain using \cref{eqn:cost_frac}:
\begin{align*}
C_{H, \nu} &=  \frac{a^2}{\alpha} + \frac{b^2}{1-\alpha}\\
& = \big(\cosh(\nu)\cosh[2](\left(1- \gamma\right) \nu) \\
&\quad + \sinh(\nu)\sinh[2](\left(1- \gamma\right) \nu)\big) e^{\nu}.
\end{align*}
In this Article, we will use its \emph{upper bound} as the sampling cost instead for a simpler expression:
\begin{align*}
C_{H, \nu} & \leq \big(\cosh[2](\left(1- \gamma\right) \nu) + \sinh[2](\left(1- \gamma\right) \nu)\big) \cosh(\nu)e^{\nu}\\
& = \cosh(2\left(1- \gamma\right) \nu) \cosh(\nu)e^{\nu}.
\end{align*}

\section{Comparison between Error Mitigation Techniques}\label{sec:comp_miti}
For a given error mitigation technique whose average estimation bias for a Pauli observable is $\epsilon$ and whose sampling cost is $C$, we can define $\Delta$ to be the effective observable that represents the error in each circuit run for obtaining the error-mitigated Pauli observable, which has:
\begin{align*}
\expval{\Delta} &\sim \epsilon\\
\var{\Delta} & \sim C.
\end{align*}
In another word, $\epsilon$ is the systematic bias of the error mitigation techniques while $C$ is the strength of the random errors (shot noise) of the technique.

We will label the average error after $N$ sample runs as $\overline{\Delta}$, for which we have:
\begin{align*}
\expval{\overline{\Delta}} &\sim \epsilon\\
\var{\overline{\Delta}} & \sim \frac{C}{N}.
\end{align*}
Hence, the expected square errors of applying the error mitigation technique with $N$ samples is thus:
\begin{align*}
\expval{\overline{\Delta}^2} & = \var{\overline{\Delta}} + \expval{\overline{\Delta}}^2\\
& \sim \frac{C}{N} + \epsilon^2
\end{align*}
Now suppose we have two different error mitigation techniques, and w.l.o.g. we will assume technique $1$ will have lower systematic bias $\epsilon_{1} \leq \epsilon_{2}$. If $C_1 < C_2$, then technique $1$ is obviously the better technique. If $C_2 \leq C_1$ however, there is a break-even point which both methods have similar mean square errors:
\begin{align*}
\expval{\overline{\Delta}_1^2} &\sim \expval{\overline{\Delta}_2^2}\\
N^* &\sim \frac{C_1 - C_2}{\epsilon_{2}^2 - \epsilon_{1}^2} \approx \frac{C_1}{\epsilon_{2}^2} + \mathcal{O}\left(\frac{C_2}{C_1}\right) + \mathcal{O}\left(\frac{\epsilon_{1}^2}{\epsilon_{2}^2}\right).
\end{align*}
i.e. $N^*$ is roughly the number of samples needed using technique $1$ to reach a shot noise level that is equal to the systematic bias of technique $2$ ($\epsilon_{2}$). When the number of samples $N \lesssim N^*$, the shot noise will dominate and thus we will choose technique 2 over technique 1 due to the lower sampling cost. When the number of samples increase and reach $N \gtrsim N^*$, then the systematic bias will dominate over shot noise and thus we will choose technique $1$ over technique $2$ due to the lower systematic bias. 

\section{Numerical result for another set of random parameters}\label{sec:numeric_2nd}
\begin{figure}[htbp]
    \centering
    \subfloat[]{\includegraphics[width = 0.45\textwidth]{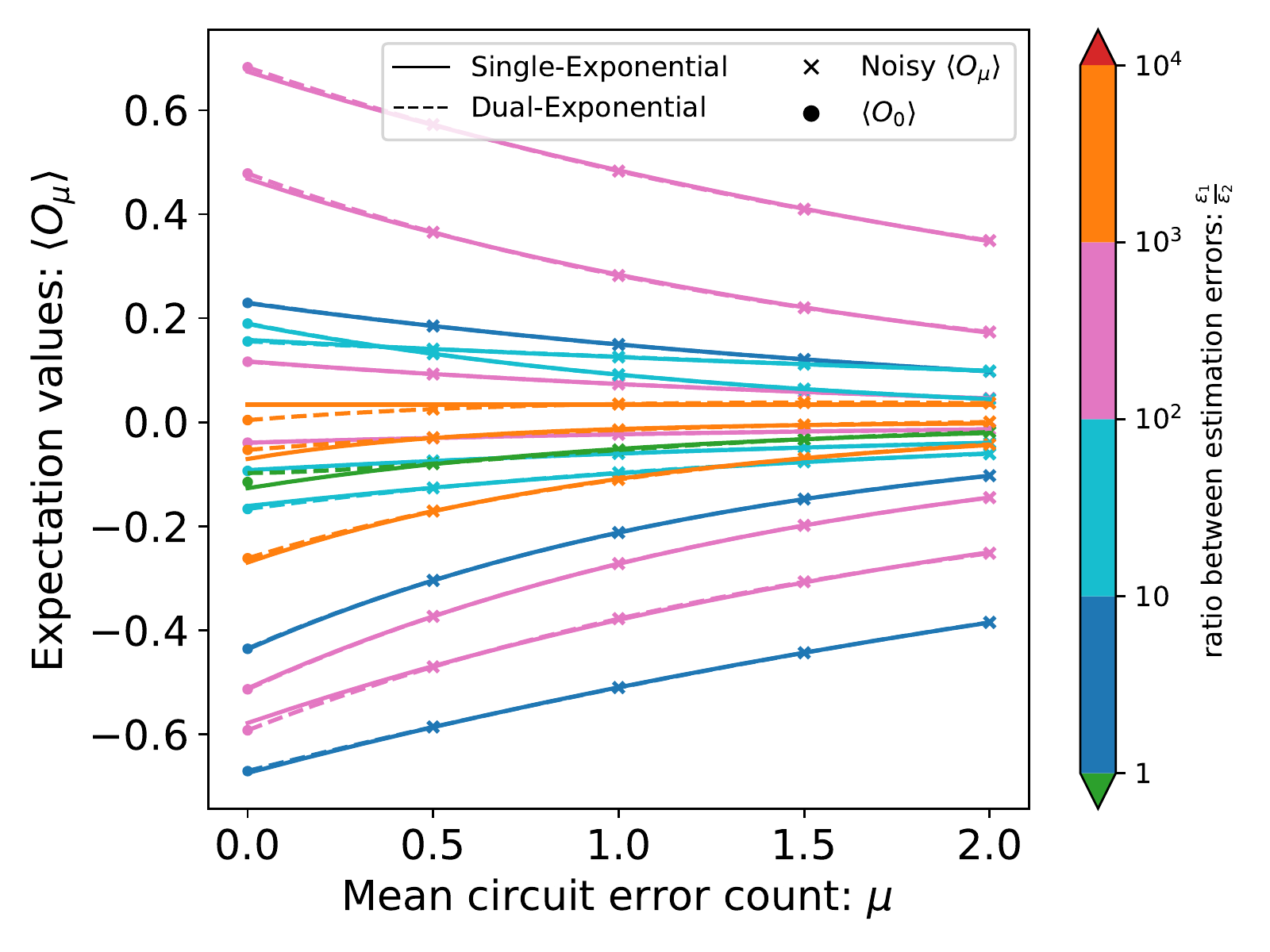}}
    \\
    \subfloat[]{\includegraphics[width = 0.45\textwidth]{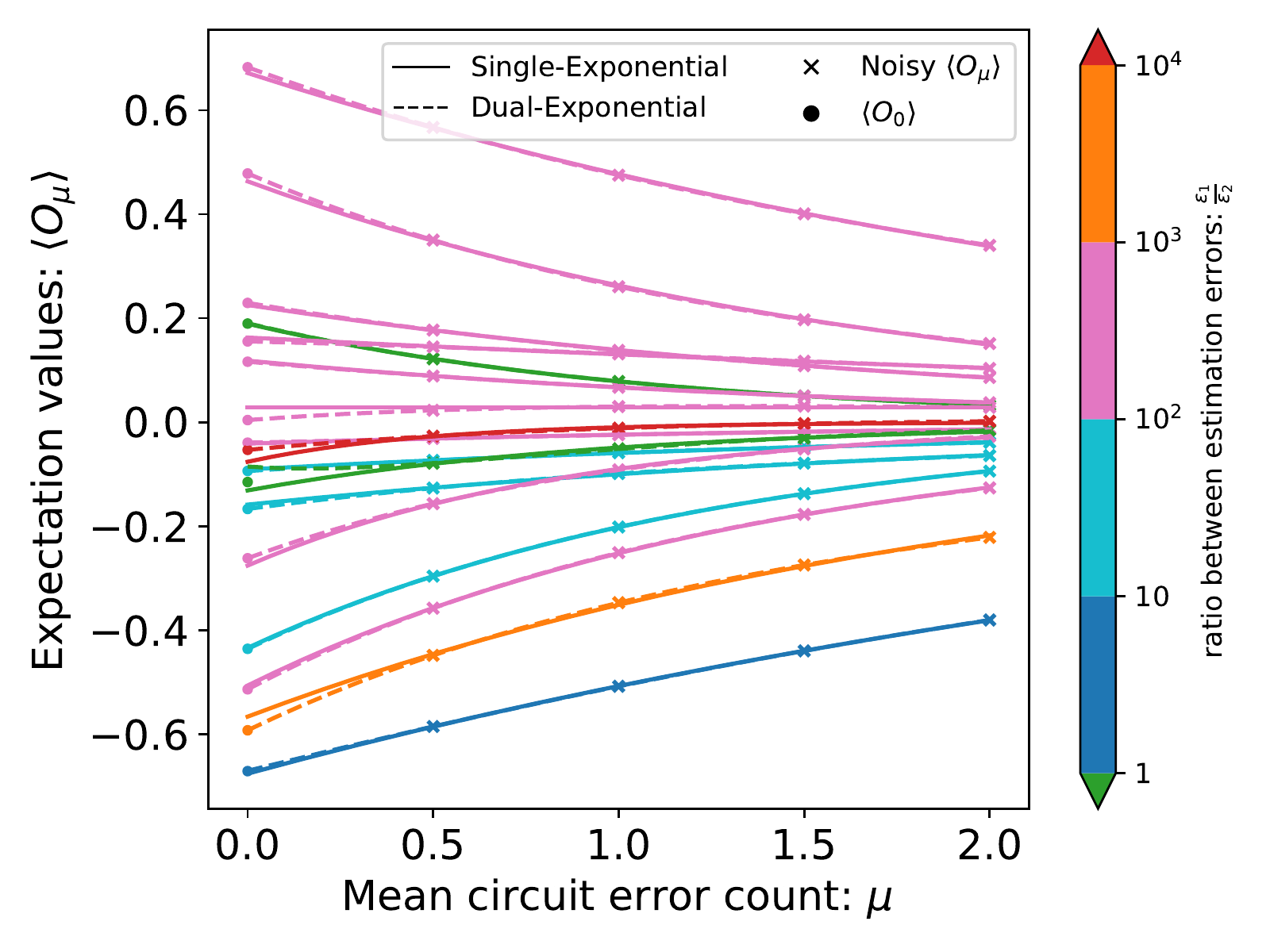}}
    \caption{Comparison between single-exponential extrapolation and dual-exponential extrapolation under (a) depolarising noise and (b) detectable noise in a $8$-qubit simulation. Plots showing the noisy expectation values of different Pauli observables obtained at the four mean circuit error counts $\mu = 0.5,\ 1,\ 1.5 ,\ 2$ (cross markers). The single- and dual-exponential extrapolation curves fitted to the data points are represented by the solid and dashed lines, respectively. The circular markers lie at $\mu = 0$ and denote the true noiseless expectation values. Different colours represent different ratios between the estimation bias of using single- and dual-exponential extrapolation}
    \label{fig:one_exp_vs_two_exp_2}
\end{figure}

\begin{figure}[htbp]
    \centering
    \subfloat[]{\includegraphics[width = 0.45\textwidth]{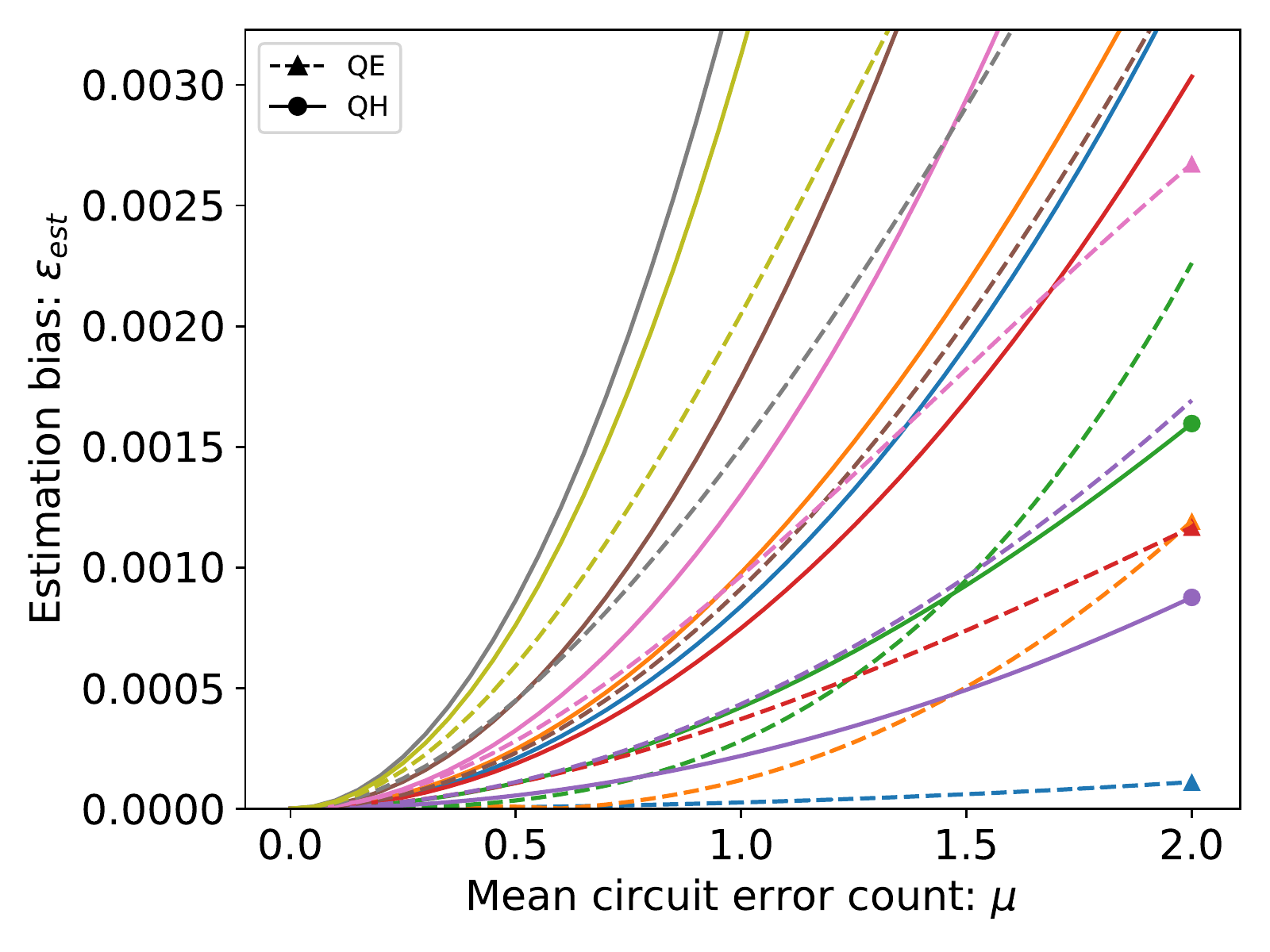}}
    \\
    \subfloat[]{\includegraphics[width = 0.45\textwidth]{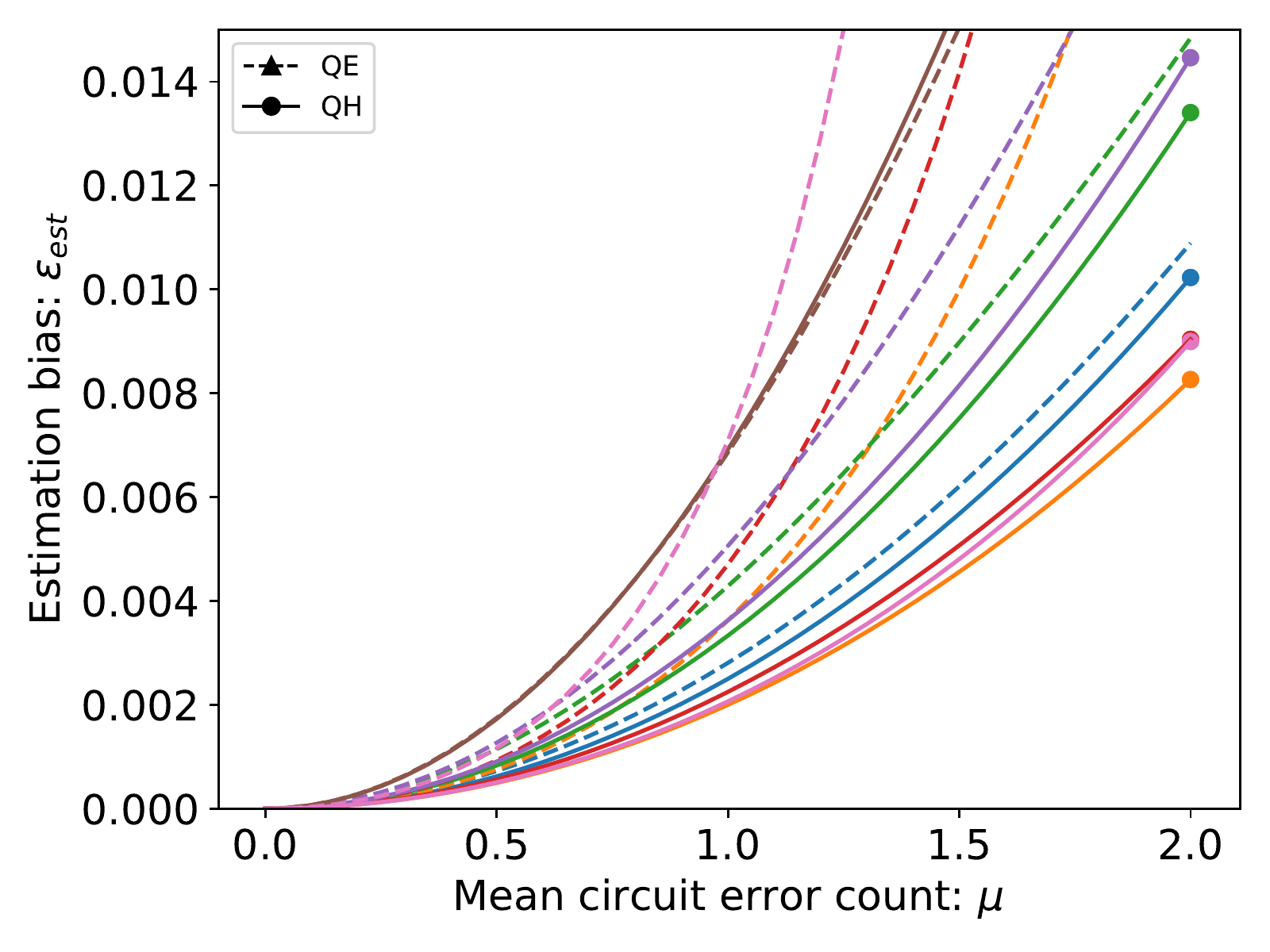}}
    \caption{Comparison of the biases in the error-mitigated expectation values between quasi-probability with exponential extrapolation (QE) and quasi-probability with hyperbolic extrapolation (QH) for (a) observables following single-exponential decay and (b) observables following dual-exponential decay in a 8-qubit simulation. Within each plot, different colours represent different observables. The solid lines denote QH, while the dashed lines denote QE. At the mean circuit error counts $\mu = 2$, for each observable, we use markers to denote the method that has lower estimation bias out of the two. For a given observable, circular markers denote lower estimation bias when using QH, while triangle markers denote lower estimation bias when using QE.}
    \label{fig:est_err_2}
\end{figure}

Here we repeat the numerical simulation in \cref{sec:expec_decay_sim} and \cref{sec:err_exp_hyp}, but with a different set of random circuit parameters. \cref{fig:one_exp_vs_two_exp_2} and \cref{tab:two_exp_err_2} show the comparison between single-exponential extrapolation and dual-exponential extrapolation, which follow similar trends as we have discussed in \cref{sec:expec_decay_sim}. \cref{fig:est_err_2} and \cref{tab:est_err_2} show the comparison between quasi-probability with exponential extrapolation and quasi-probability with hyperbolic extrapolation, which follow similar trends as we have discussed in \cref{sec:err_exp_hyp}.

\begin{table}[htbp!]
    \centering
    \begingroup
    \renewcommand{\arraystretch}{1.5}
    \begin{tabular}{lcc}\toprule
        $\overline{\epsilon}_{1}, \overline{\epsilon}_{2} / 10^{-4}$ &Depolarising  & Detectable\\ \hline
        Single-exp& 67 &96  \\
        Dual-exp& 1.1& 1.3\\
        \botrule
    \end{tabular}
    \endgroup
    \caption{The bias in the single- and dual-extrapolation estimates averaged over observables within each plots in \cref{fig:one_exp_vs_two_exp_2} excluding observables with exceptionally large $\epsilon_{1}$ or $\epsilon_{2}$. The entries are in the unit of $10^{-4}$. }
    \label{tab:two_exp_err_2}
\end{table}

\begin{table}[htbp!]
    \centering
    \begingroup
    \renewcommand{\arraystretch}{1.5}
    \subfloat[$\mu=1$ ]{
        \begin{tabular}{cccc}\toprule
            $\overline{\epsilon}_{est}/ 10^{-4}$& \  1-Exp. Obs. \  & \  2-Exp. Obs. \ & \  All Obs. \ \\ \hline
            QE &7.4&49&26\\
            QH &14&32&22\\
            \botrule
        \end{tabular}
    }\\
    \subfloat[$\mu=2$ ]{
        \begin{tabular}{cccc}\toprule
            $\overline{\epsilon}_{est}/ 10^{-3}$& \  1-Exp. Obs. \  & \  2-Exp. Obs. \ & \  All Obs. \ \\ \hline
            QE &2.5&110&49\\
            QH &6.1&13&9.2\\
            \botrule
        \end{tabular}
    }
    
    \endgroup
    \caption{The biases in the error-mitigated estimates using QE and QH averaged over single-exponential, dual-exponential and all observables at the mean circuit error counts (a) $\mu = 1$ and (b) $\mu = 2$. The entries in (a) and (b) are in the units of $10^{-4}$ and $10^{-3}$, respectively.}
    \label{tab:est_err_2}
\end{table}
\FloatBarrier
\newpage
%

\end{document}